\documentclass{emulateapj}
\usepackage{epsfig,graphics}
\usepackage[colorlinks=true,  citecolor=blue,  pagecolor=blue,
  linkcolor=blue,  menucolor=blue, urlcolor=blue,
  linkbordercolor={0 0 1}, frenchlinks=True,  breaklinks]{hyperref}

\begin{document}
\newcommand{\Ha}{H$\alpha$}
\newcommand{\Hb}{H$\beta$}
\newcommand{\NII}{[N \textsc{ii}]}
\newcommand{\OIII}{[O \textsc{iii}]}
\newcommand{\OII}{[O \textsc{ii}]}
\newcommand{\zphotf}{z_{\rm phot}}
\newcommand{\zspecf}{z_{\rm spec}}
\newcommand{\zphot}{$\zphotf$}
\newcommand{\zspec}{$\zspecf$}
\newcommand{\Rcf}{R_{\rm C}}
\newcommand{\Rc}{$\Rcf$}

\newcommand{\dz}{$\Delta z_{68}$}
\newcommand{\galex}{{\it GALEX}}
\newcommand{\spitzer}{{\it Spitzer}}
\newcommand{\nuv}{{\it NUV}}
\newcommand{\fuv}{{\it FUV}}
\newcommand{\EBV}{E(B-V)}

\newcommand{\Lya}{Ly$\alpha$}
\newcommand{\mm}{$\mu$m}
\newcommand{\Msun}{M$_{\sun}$}

\newcommand{\iyr}{yr$^{-1}$}
\newcommand{\vMpc}{Mpc$^{-3}$}
\newcommand{\sbzk}{sBzK}

\newcommand{\fBK}{5BK3$\sigma$}
\newcommand{\fBf}{5B5$\sigma$}

\newcommand{\Un}{U_n}
\newcommand{\Rs}{\mathcal{R}}
\newcommand{\UBV}{U -- BV}
\newcommand{\BVRi}{BV -- \Rc i\arcmin}

\newcommand{\rNB}{{\rm NB}}

\newcommand{\iz}{$i$\arcmin--$z$\arcmin}

\newcommand{\zNB}{$z$\arcmin--\rNB}
\newcommand{\zNBa}{$z$\arcmin--{\rm NB921}}
\newcommand{\zNBb}{$z$\arcmin--{\rm NB973}}
\newcommand{\nOII}{$\sim$1,300}

\newcommand{\nsourcea}{119,541}
\newcommand{\nsourceb}{84,786}

\newcommand{\areaa}{870.4}
\newcommand{\areab}{788.7}

\newcommand{\za}{1.47}
\newcommand{\zb}{1.62}

\newcommand{\consta}{-0.003}
\newcommand{\slopea}{0.077}
\newcommand{\constb}{-0.014}
\newcommand{\slopeb}{0.271}

\newcommand{\Nema}{2512} 
\newcommand{\Nemaf}{2361}
\newcommand{\Nemb}{1666} 
\newcommand{\Nembf}{1243}

\newcommand{\Nemza}{2261}  
\newcommand{\Nemzaf}{2132} 
\newcommand{\Nemzb}{1403}  
\newcommand{\Nemzbf}{1040} 

\newcommand{\speca}{522}  
\newcommand{\specb}{204}  

\newcommand{\specga}{295} 
\newcommand{\specgb}{104} 
\newcommand{\specua}{31}  
\newcommand{\specub}{8}   

\newcommand{\specHaa}{81}
\newcommand{\specOIIIa}{80}
\newcommand{\specOIIa}{104}
\newcommand{\specLyaa}{21}

\newcommand{\specHab}{49}

\newcommand{\specOIIIb}{12}
\newcommand{\specOIIb}{32}
\newcommand{\specLyab}{1}

\newcommand{\Nemaz}{2013}
\newcommand{\Nembz}{1131}

\newcommand{\Ndual}{241}
\newcommand{\Ndualz}{80}
\newcommand{\Ndualbad}{nine}

\newcommand{\zphotmina}{1.440}
\newcommand{\zphotmaxa}{1.485}

\newcommand{\zphotminb}{1.591}
\newcommand{\zphotmaxb}{1.644}

\newcommand{\NOIIa}{933}
\newcommand{\NOIIb}{328}

\newcommand{\POIIzphota}{77\%}
\newcommand{\POIIzphotb}{58\%}

\newcommand{\NOIIaK}{750}
\newcommand{\NOIIbK}{264}

\newcommand{\NOIIaKdet}{402}
\newcommand{\NOIIbKdet}{174}

\newcommand{\sbzka}{338}
\newcommand{\sbzkap}{84\%}

\newcommand{\sbzkb}{158}
\newcommand{\sbzkbp}{91\%}

\newcommand{\SFRslope}{0.89\pm0.22}
\newcommand{\SFRint}{16.23}
\submitted{Received 2012 May 31}
\title{The Stellar Population and Star Formation Rates of $z\approx1.5$--1.6
  \OII\ Emitting Galaxies Selected from Narrow-Band Emission-Line Surveys}
\author{Chun Ly,\altaffilmark{1,9}
  Matthew A. Malkan,\altaffilmark{2}
  Nobunari Kashikawa,\altaffilmark{3,4}
  Masao Hayashi,\altaffilmark{3}
  Tohru Nagao,\altaffilmark{5,6}
  Kazuhiro Shimasaku,\altaffilmark{7,8}
  Kazuaki Ota,\altaffilmark{6} and 
  Nathaniel R. Ross\altaffilmark{2}}

\shorttitle{Stellar Population and SFRs of \OII\ Star-Forming Galaxies}
\shortauthors{Ly et al.}
\email{chunly@stsci.edu}
\altaffiltext{1}{Space Telescope Science Institute, Baltimore, MD, USA}
\altaffiltext{2}{Department of Physics and Astronomy, UCLA, Los Angeles, CA, USA}
\altaffiltext{3}{Optical and Infrared Astronomy Division, National Astronomical
  Observatory, Mitaka, Tokyo, Japan}
\altaffiltext{4}{Department of Astronomy, School of Science, Graduate University
  for Advanced Studies, Mitaka, Tokyo, Japan}
\altaffiltext{5}{The Hakubi Project, Kyoto University, Kyoto, Japan }
\altaffiltext{6}{Department of Astronomy, Kyoto University, Kyoto, Japan}
\altaffiltext{7}{Department of Astronomy, School of Science, University of Tokyo,
  Bunkyo, Tokyo, Japan}
\altaffiltext{8}{Research Center for the Early Universe, School of Science,
  University of Tokyo, Tokyo, Japan}
\altaffiltext{9}{Giacconi Fellow.}

\begin{abstract}
  We present the first detailed study of the stellar populations of 
  star-forming galaxies at $z\sim1.5$, which are selected by their
  \OII\ emission line, detected in narrow-band surveys.
  We identified \nOII\ \OII\ emitters at $z=\za$ and $z=\zb$ in the
  Subaru Deep Field with rest-frame EWs above 13\AA. 
  Optical and near-infrared spectroscopic observations for
  $\approx$10\% of our samples show that our separation of \OII\
  from \OIII\ emission-line galaxies in two-color space is 99\%
  successful.
  We analyze the multi-wavelength properties of a subset of $\sim$1,200
  galaxies with the best photometry. They have average rest-frame EW
  of 45\AA, stellar mass of $3\times10^{9}$ \Msun, and stellar age of
  100 Myr. In addition, our SED fitting and broad-band colors indicate
  that \OII\ emitters span the full range of galaxy populations at
  $z\sim1.5$.
  We also find that 80\% of \OII\ emitters are also photometrically
  classified as ``BX/BM'' (UV) galaxies and/or the star-forming
  ``BzK'' (near-IR) galaxies. Our \OII\ emission line survey produces
  a far more complete, and somewhat deeper sample of $z\sim1.5$
  galaxies than either the BX/BM or sBzK selection alone.
  We constructed average SEDs and find that higher \OII\ EW galaxies
  have somewhat bluer continua. SED model-fitting shows that they
  have on average half the stellar mass of galaxies with lower \OII\ EW.
  The observed \OII\ luminosity is well-correlated with the far-UV
  continuum with a logarithmic slope of $\SFRslope$.
  The scatter of the \OII\ luminosity against the far-UV continuum
  suggests that \OII\ can be used as a SFR indicator with a reliability
  of 0.23 dex.
\end{abstract}


\defcitealias{ly07}{L07}
\defcitealias{salpeter}{Salpeter}


\section{INTRODUCTION}\label{1}
Since the star formation rate (SFR) density  of galaxies is an order of
magnitude higher at $z\approx1$--5 than in the local universe
\citep[e.g.,][]{hopkins04}, emission-line galaxies are prominent in the
young universe, and thus are useful probes of the evolution of galaxies.
Current techniques to identify such emission-line galaxies include
grism surveys \citep{shim09,atek10,coil11,straughn11}, slit spectroscopy
\citep[e.g., DEEP2;][]{davis03}, and narrow-band imaging
\citep[e.g.,][hereafter L07]{malkan96,teplitz98,fujita03,hippelein03,ly07}.
Narrow-band surveys have identified large samples of star-forming galaxies
by detecting their redshifted nebular emission lines, at low and intermediate
redshifts, to as high as $z\sim7$. These surveys have the ability to:
(1) determine redshift to 1 percent accuracy, (2) derive emission-line
fluxes, which can be used to determine star-formation rates (SFRs) and
trace the evolution of the cosmic SFR density. And they accomplish this with
roughly an order-of-magnitude higher efficiency than spectroscopic surveys.

However, it is still unclear what galaxy population(s) these techniques
probe or are unable to probe. For example, it might be that these surveys
are biased towards the youngest and possibly least reddened galaxies.
Such claims have been made for Ly$\alpha$ emitters when compared to
Lyman break galaxies at similar redshifts \citep[e.g.,][]{gawiser07}.
Thus, it is crucial to understand the selection bias of current and future
emission-line surveys using these techniques.
Our survey utilizes \nOII\ \OII\ emitters at $z=\za$ and $z=\zb$
in the Subaru Deep Field \citep[SDF;][]{kashik04}.

The outline of this paper is as follows. In Section~\ref{2}, we discuss the
SDF and the optical imaging that we have acquired and reduced.
Section~\ref{sec:selection} discusses the selection of narrow-band
(NB) excess emitters, the spectroscopic observations, the technique to
separate \OII\ emitters from \Ha\ and \OIII, and the photometric redshifts.
Section~\ref{4} presents the results on the \OII\ EW distributions,
a comparison between the rest-frame UV luminosity and the \OII\ luminosity,
their stellar population from modeling the spectral energy distribution (SED),
and compares the \OII\ population against popular color selection techniques
at these redshifts. We then summarize our conclusions in Section~\ref{sec:5}.
Throughout this paper, we adopt a flat cosmology with $\Omega_{\Lambda}=0.7$,
$\Omega_M=0.3$, and $H_0=70$ km s$^{-1}$ Mpc$^{-1}$, and magnitudes are
reported on the AB system \citep{oke74}.


\section{The Subaru Deep Field Observations}\label{2}
Several extragalactic fields already have deep imaging in NB filters with
Subaru/Suprime-Cam \citep{miyazaki02}. 
Of all of these, the SDF has the most sensitive optical NB imaging in six
NB filters in the sky, and is further complemented with deep multi-band
imaging between 1500\AA\ and 4.5\mm.
The most prominent emission lines entering these NB filters are \Ha,
\OIII\ $\lambda$5007, and \OII\ $\lambda\lambda$3726,3729, at
well-defined redshift windows between $z\approx0.07$ and $z\approx\zb$.
%
%
\subsection{NB921 and NB973 Imaging Observations}
For this paper, we focus on excess emitters found using two
filters, NB921 and NB973. Their central wavelengths and FWHM
are 9196\AA\ and 132\AA\ for NB921 and 9755\AA\ and 200\AA\ for NB973.
The acquisition of the SDF NB921 data has already been discussed in
\cite{kashik04}, \cite{kashik06}, and \citetalias{ly07}.
The NB921 data are identical to the publicly released SDF v.1
data.\footnotemark\
The NB973 data were acquired on 2005 March 16--17 and 2007 May 9 for
a total integration time of 21.5 h \citep{iye06,ota08}. The seeing varied
between 0\farcs5 and 1\farcs0\ and observing conditions were photometric
in 2005 but not in 2007.
\footnotetext[10]{\url{http://soaps.nao.ac.jp/SDF/v1/index.html}.}

Both the NB921 and NB973 data were reduced using \textsc{sdfred}
\citep{yagi02,ouchi04}, a software package designed solely for
Suprime-Cam data.
The spatial resolutions for the final NB921 and NB973 mosaics are
0\farcs96 and 0\farcs91, respectively.

By repeatedly measuring the flux within 2\arcsec\ diameter
apertures at random positions in the mosaics, we determined that the
3$\sigma$ limits for the NB921 and NB973 data are 26.60 and 25.64 mag,
respectively. A summary of the sensitivity in other wave-bands,
that are later discussed, can be found in Table 1 of \cite{ly11a}.
%
The publicly available $z$\arcmin-band mosaicked image, which is used
throughout this paper, has a 3$\sigma$ sensitivity of 26.83 mag,
while the $i$\arcmin-band data (later used) is 0.93 mag deeper.


\section{Selection of Narrow-Band Excess Emitters}\label{sec:selection}
The NB921- and NB973-detected catalogs were obtained by running
SExtractor \citep[vers. 2.5.0]{bertin96}.
The unmasked regions cover \areaa\ arcmin$^2$ for NB921 and
\areab\ arcmin$^2$ for NB973. The latter is smaller due to higher
systematic noise in one of the ten CCDs, which was masked
to avoid significant spurious detections.
A total of \nsourcea\ and \nsourceb\ sources were detected in the
unmasked regions of the NB921 and NB973 images, respectively.

To select NB excess emitters, the standard technique is to compare
the flux in the NB against the flux in an adjacent broad-band filter.
We use the $z$\arcmin-band, since it samples the local continuum around
these NB filters. Fluxes from the $z$\arcmin-band were extracted by
running SExtractor in ``dual-image'' mode, where the respective NB image
was used as the ``detection'' image.
This works well because all three mosaicked images have very similar
seeing, so the \zNB\ colors are determined within the same physical
scale of the galaxies.
For the extraction of fluxes and selection of sources, we use a 2\arcsec\
diameter circular aperture, which encloses 17 kpc at $z\approx1.5$.
%
%
\begin{figure}
  \epsscale{1.1}
  \plotone{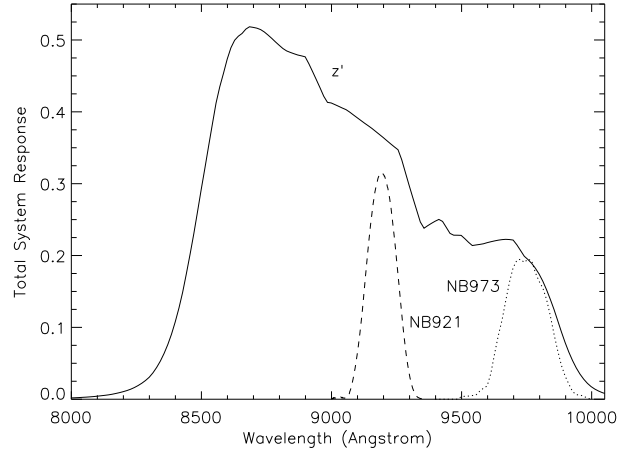}
  \caption{Total system throughput for the $z$\arcmin\ (solid),
    NB921 (dashed), and NB973 (dotted) filters.}
  \label{response}
\end{figure}

One minor issue affecting the NB921 and NB973 filters is that they
are located redward of the $z$\arcmin-band sensitivity peak,
$\sim$8700\AA\ (see Figure~\ref{response}). For redder
galaxies, this makes the \zNB\ colors systematically higher, and
could falsely identify lineless red galaxies as NB-excess emitters.

\begin{figure*}
  \epsscale{1.1}
  \plottwo{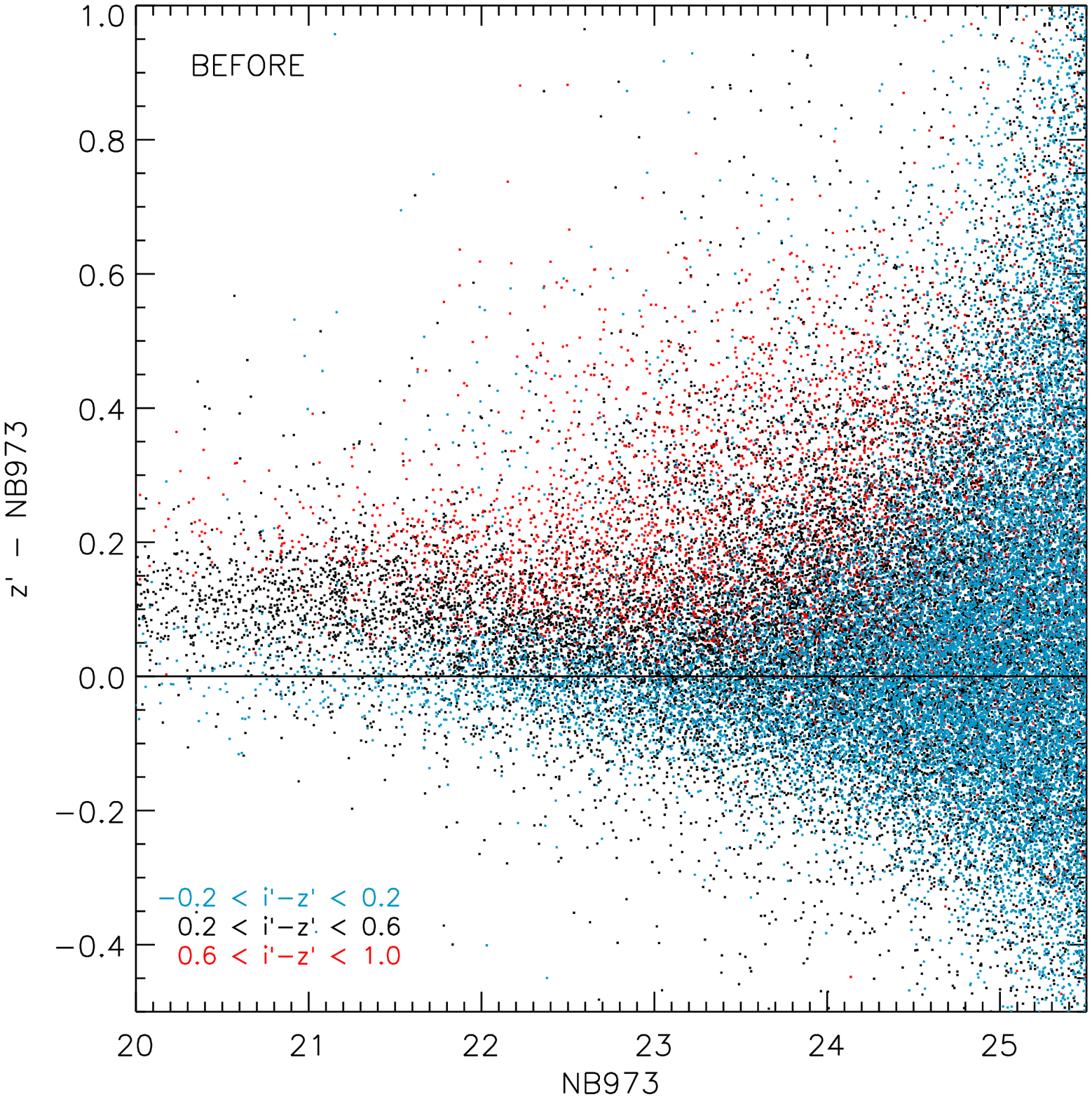}{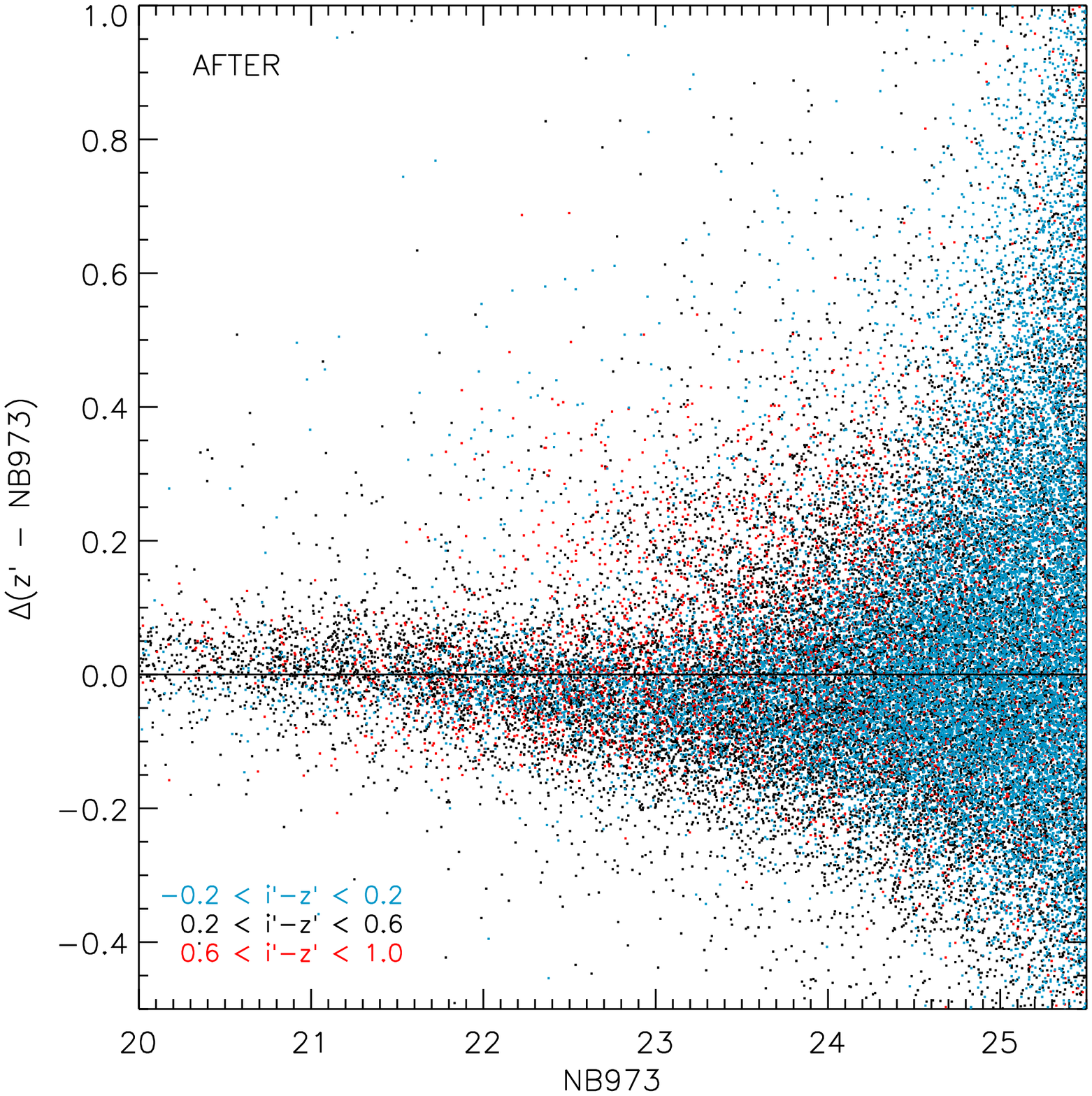}
  \caption{$z$\arcmin--NB973 colors for NB973 excess emitters as
      measured (left) and after applying the \iz\ color-term correction
      given in Equation~(\ref{ctb}) (right). Cyan, black, and red squares
      denote galaxies with blue, intermediate, and red \iz\ colors.
      The color-term correction ensures that red galaxies with weak
      or absent \OII\ emission are not falsely classified as NB973
      excess emitters.}
  \label{iz_corr}
\end{figure*}

To resolve this issue, we empirically estimate the continuum flux at
the wavelengths of the NB filters by fitting the
\iz\ and \zNB\ colors for non-NB excess sources:
\begin{eqnarray}
  \label{cta}
  \langle z\arcmin-{\rm NB921}\rangle_{\rm C} &=& \consta\ + \slopea(i\arcmin-z\arcmin),{\rm~and}\\
  \label{ctb}
  \langle z\arcmin-{\rm NB973}\rangle_{\rm C} &=& \constb\ + \slopeb(i\arcmin-z\arcmin).
\end{eqnarray}
These relations are consistent with simple linear extrapolations of
the \iz\ color to 9200\AA\ or 9755\AA. The corrected \zNB\ color is then
\begin{equation}
  \Delta(z\arcmin-{\rm NB})=(z\arcmin-{\rm NB}) - \langle z\arcmin - {\rm NB}\rangle_{\rm C}.
\end{equation}
To illustrate the reliability of these corrections, we plot the
$z$\arcmin--NB973 colors with and without the color-term correction for
NB973 emitters in Figure~\ref{iz_corr}.
%
%
This correction was not performed previously for the NB921 excess
emitter sample identified by \citetalias{ly07}.
It has a minor effect for the NB921 sample, but is more
important for the NB973 sample since the NB973 filter is
significantly redder by an additional $\sim$560\AA.
For example, we find that the rms in the \zNB\ color (at the bright end)
is decreased by 10\% for NB921 and 50\% for NB973 when this correction
is applied (see Figure~\ref{iz_corr}).
%
Since the minimum threshold for selection (discussed below) is driven
by the observed scatter, such a reduction permits us to select toward
lower minimum \zNB\ excesses.
%
%
\begin{figure*}
  \epsscale{1.1}
  \plottwo{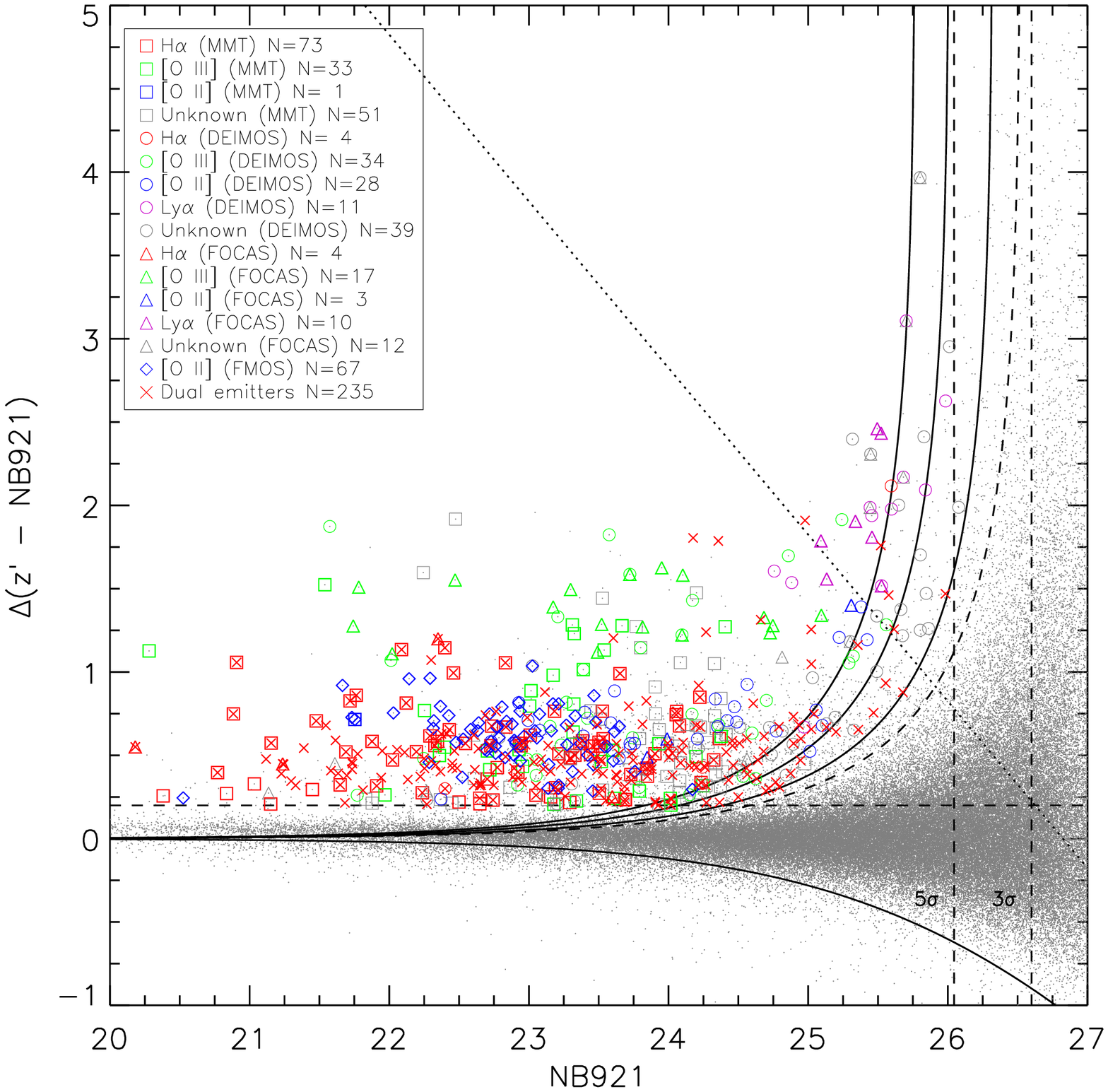}{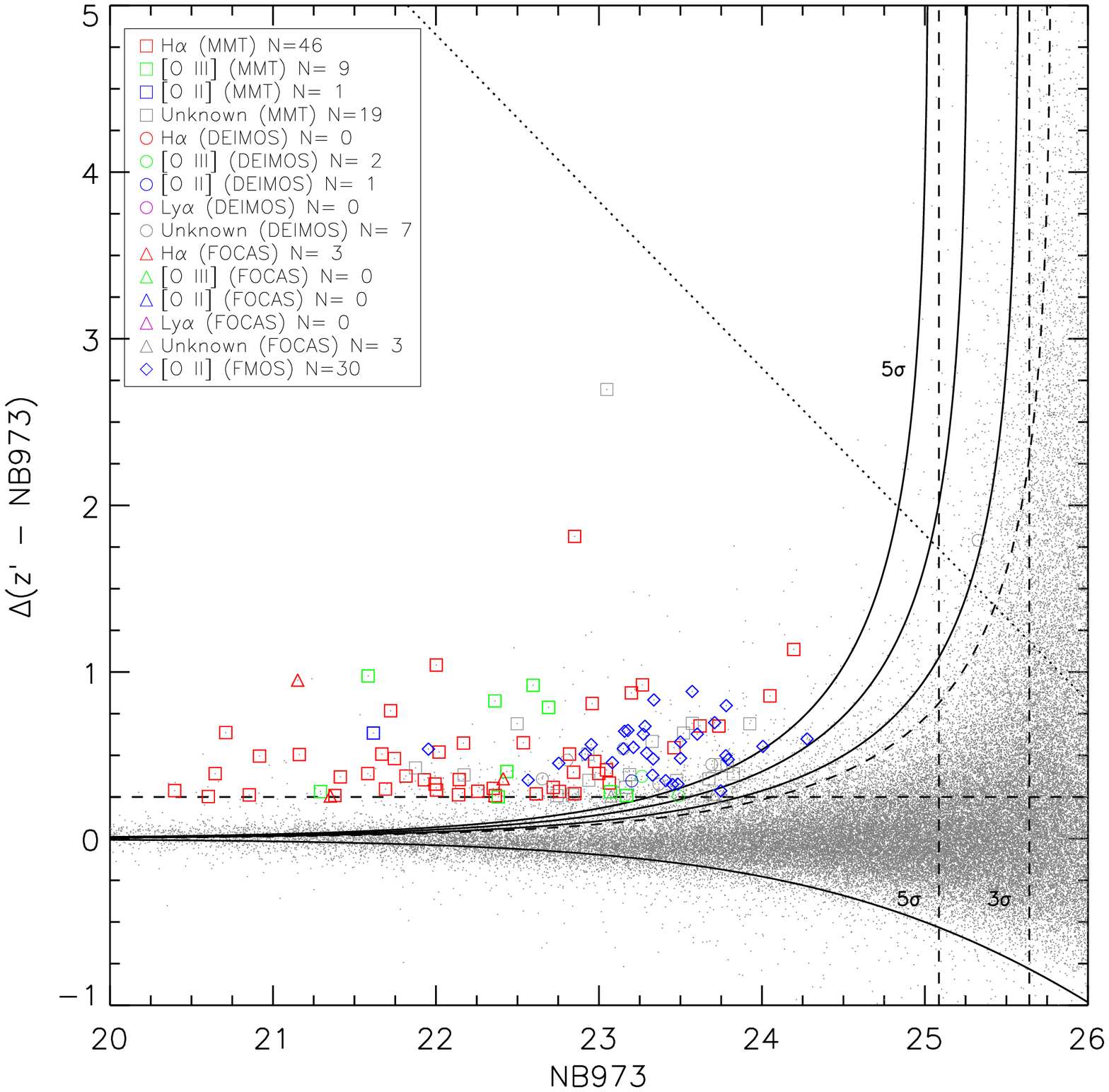}
  \caption{$\Delta$(\zNB) for NB921 (left) and NB973 (right) samples
    (grey points) in the SDF. The $3\sigma$ and $5\sigma$ sensitivity
    limits for the NB data are shown as
    vertical dashed lines. Minimum excesses of 0.2 mag (left) and 0.25
    mag (right) are illustrated with the horizontal dashed lines. Also
    overlaid are curves for $N\times\sigma_{z\arcmin-\rNB}$ with $N=2.5$
    (dashed line) and 3.0, 4.0, and 5.0 (solid lines).
    Sources that fall above the diagonal dotted lines are undetected
    in the $z$\arcmin-band at 3$\sigma$. In the absence of noise, 
    the strongest possible magnitude excess is
    $\approx$2.5 magnitudes, which corresponds to a pure emission line
    centered in the NB filter, and no continuum.
    Spectroscopically confirmed sources are overlaid with symbols
    representing different instruments: DEIMOS (circles), Hectospec
    (squares), FOCAS (triangles), and FMOS (diamonds).
    Red crosses represent galaxies that are dual excess emitters
    in both the NB704 and NB921 bands (see Section~\ref{dual_lines}).
    \Ha, \OIII, \OII, and Ly$\alpha$ are distinguished by red, green,
    blue, and purple colors, respectively. Also overlaid in grey
    symbols are the cases where an emission line was not detected
    in the spectra.}
  \label{zNB}
\end{figure*}

In Figure~\ref{zNB}, we illustrate the corrected \zNBa\ and \zNBb\
colors for NB921- and NB973-detected sources, respectively. For
selecting objects with genuine  NB921 and NB973 excesses, we
require minimum \zNB\ magnitude differences of 0.20 mag
(observed EW of 32\AA) and 0.25 mag (observed EW of 70\AA),
respectively.
These minima are 5.6 and 6.7 times the rms of the \zNB\ color at
the luminous end.
The observed EW is determined from a combination of the broad-
and narrow-band fluxes, and is given by:
\begin{equation}
  {\rm EW}_{\rm obs}=\Delta\rNB \left[\frac{f_{\lambda,\rNB}-f_{\lambda,z}}{f_{\lambda,z}-f_{\lambda,\rNB}(\Delta{\rNB}/\Delta{z})}\right],
\end{equation}
where $\Delta{\rm NB}$ and $\Delta z$ (955\AA) are the FWHMs for
the narrow-band and $z$\arcmin, respectively, and
$f_{\lambda,X}$ is the flux density in
erg s$^{-1}$ cm$^{-2}$ \AA$^{-1}$ in band ``$X$'' \citepalias{ly07}.
We also require that the \zNB\ color exceeds
\begin{equation}
\nonumber
3\sigma_{z\arcmin-\rNB}(\rNB) \equiv -2.5\log\left(1-\frac{\sqrt{f_{\nu,3\sigma\rNB}^2+f_{\nu,3\sigma z\arcmin}^2}}{f_{\nu,\rNB}}\right),
\end{equation}
where $f_{\nu,3\sigma X}$ is the 3$\sigma$ flux density limit
for band ``$X$'' and $f_{\nu,\rNB}$ is the flux density for a given
NB magnitude.
These selection criteria yield \Nema\ NB921 excess emitters and \Nemb\ NB973
excess emitters. The 3$\sigma$ emission-line flux limits are
$5.8\times10^{-18}$ and $1.7\times10^{-17}$ erg s$^{-1}$ cm$^{-2}$
for NB921 and NB973, respectively.
Smaller samples were also constructed by requiring stronger detections
of 4$\sigma_{z\arcmin-\rNB}$ and 5$\sigma_{z\arcmin-\rNB}$.
These samples contain 1912 and 1616 for NB921 and 839 and 565 for NB973.

While the color-term corrections (see Equations~(\ref{cta})--(\ref{ctb}))
should apply to galaxies with SEDs that behave simply, at
$z\sim1.3$--1.4 the 4000\AA\ break falls in the $z$\arcmin-band, but
is blueward of the NB filter. This poses a problem, as galaxies with
older stellar populations ($\gtrsim$1 Gyr) will have a strong break,
and the continuum flux at $\sim$9200\AA\ or 9700\AA\ can easily be
underestimated from a linear extrapolation of the \iz\ color. This
population was confirmed when the photometric redshifts (hereafter
photo-$z$'s or \zphot; see Section~\ref{sec:photoz} for further
discussion) were derived. At these \zphot's,
the \OII\ emission line does not fall into the NB973 passband.
Since these galaxies are very red, simply imposing a cut
of \iz\ $\leq$ 0.8 (0.4) for NB921 (NB973), most of these 4000\AA-break
galaxies can be removed without affecting the true line emitters.
By removing these red false NB excess emitters, the NB921 sample is
reduced to \Nemaf\ (3$\sigma$), 1787 (4$\sigma$), and 1505 (5$\sigma$). 
Likewise, the NB973 sample is reduced by about 25\%--30\%, to \Nembf\ (3$\sigma$),
571 (4$\sigma$), and 375 (5$\sigma$).

\subsection{Independent Confirmation of NB Excesses}

\subsubsection{Follow-up Spectroscopy of NB Excess Candidates}
\label{sec:spec}
To confirm that the NB technique efficiently identifies emission-line
galaxies, we have conducted spectroscopy with several instruments
(both optical and near-infrared) over the past several years.
The observations were obtained with Keck's Deep Imaging
Multi-Object Spectrograph \citep[DEIMOS;][]{faber03}
in 2004, 2008, and 2009, Subaru's Faint Object Camera
and Spectrograph \citep[FOCAS;][]{kashik02} in 2004 and 2007,
MMT's Hectospec \citep{fabricant05} in 2008, and Subaru's Fiber
Multi-object Spectrograph \citep[FMOS;][]{kimura10} in 2012.

The DEIMOS and FOCAS follow-up spectroscopy has been discussed in
\cite{kashik06} and \citetalias{ly07}, and DEIMOS data that were
acquired more recently are discussed in \cite{kashik11}.
The Hectospec and FOCAS observations typically had 1--2 hrs
of on-source integration, while the DEIMOS observations had
2--4 hrs. In addition, we observed for 2.5--3.0 hrs with FMOS.

For the Hectospec observations, we utilized the 270 mm$^{-1}$
grating blazed at 5200\AA\ to yield a spectral coverage of
3650\AA--9200\AA\ with a spectral resolution of $\approx$5\AA.
A total of four different fiber configurations were used.
The on-source integration varied between 80 and 140 minutes.
The MMT spectra were reduced following standard procedures with the
\textsc{hsred}\footnotemark\ reduction pipeline.
\footnotetext[11]{\url{http://www.astro.princeton.edu/\~rcool/hsred/}.}

FMOS observations, which only targeted galaxies suspected to be
\OII\ emitters (see Section~\ref{OIIid}), were obtained on 2012 8--9 Apr utilizing
``cross-beam switching'' mode. Each exposure was 15 m in length
and obtained with ``ABAB'' dithering. The data were reduced
using \textsc{fibre}, a reduction package that consists of
\textsc{iraf} scripts and \textsc{cfitsio} routines,
designed solely for optimal FMOS data reduction \citep{iwamuro11}. 

\newcommand{\ps}{\phm{1}}
\begin{deluxetable*}{ccccrrrrcccc}
  \tabletypesize{\scriptsize}
  \tablewidth{0pc}
  \tablecaption{Summary of Spectroscopic Samples}
  \tablehead{
    \colhead{Filter}&
    \colhead{Targeted\tablenotemark{1}}&
    \colhead{Confirmed\tablenotemark{1}}&
    \colhead{Ambiguous\tablenotemark{1}}&
    \colhead{DEIMOS}&
    \colhead{MMT}&
    \colhead{FOCAS}&
    \colhead{FMOS\tablenotemark{2}}&
    \colhead{$N$(\Ha)}&
    \colhead{$N$(\OIII)}&
    \colhead{$N$(\OII)}&
    \colhead{$N$(\Lya)}\\
    \colhead{(1)}&\colhead{(2)}&\colhead{(3)}& \colhead{(4)}& \colhead{(5)}& \colhead{(6)}&
    \colhead{(7)}&\colhead{(8)}&\colhead{(9)}&\colhead{(10)}&\colhead{(11)}&\colhead{(12)}}
  \startdata
  NB921 &  \speca & \specga &  227 &  88 (127) & 113 (164) & 36 (48) &  68 (212) & \specHaa & \specOIIIa & \specOIIa & \specLyaa\\
  NB973 &  \specb & \specgb &  100 &   6 ( 13) &  61 ( 80) &  6 ( 9) &  30 (105) & \specHab & \specOIIIb & \specOIIb & \specLyab\\
  \vspace{-3mm}
  \enddata
  \label{tab:spec_table}
  \tablecomments{The number of targeted, spectroscopically confirmed, and
    unclassified galaxies are given in Cols. (1)--(3). Values in
    parentheses denote the total targeted sample with each instrument.
    Confirmed galaxies that are neither \Ha, \OIII, \OII, or \Lya\ are
    discussed in the text.}
  \tablenotetext{1}{Note that a subset of the samples were targeted with
    different instruments. These numbers account for overlapping targets.}
  \tablenotetext{2}{The FMOS reduction is not complete, as improvements
    for optimal spectral stacking and detection of weak emission lines
    have yet to be implemented. Thus, these confirmed numbers represent lower
    limits.}
\end{deluxetable*}

A summary of the spectroscopic observations obtained for
NB921 and NB973 emitters is provided in Table~\ref{tab:spec_table}.
In total, \speca\ NB921 emitters and \specb\ NB973 emitters
were targeted.
Our NB973 spectroscopic observations are sparser, since that
sample became available much later. 
The majority (70\% and 90\% for the NB921 and NB973
samples, respectively) of our spectra are from Hectospec
and FMOS, with most of the \OII\ confirmations from FMOS.
Among these spectroscopically targeted NB921 excess emitters,
$\approx$57\% (\specga) have reliable redshifts (based
at least one other emission line).
Similarly, 51\% (\specgb) of the observed NB973 excess
emitters yielded clear spectroscopic identifications and redshifts.
These success rates were closer to 70\% using only the optical
spectroscopy, which is generally more sensitive.
In addition, we exclude \specua\ (6\%) NB921 and \specub\ (3\%)
NB973 sources that show single weak emission lines in their optical spectra
which, if they are real, are consistent with the redshifts predicted
by an emission line in the NB filter.

The numbers of spectroscopically confirmed \Ha, \OIII/\Hb, \OII,
and Ly$\alpha$ lines are summarized in Table~\ref{tab:spec_table}.
We note that a small subset (9 NB921 and 10 NB973) of our
spectroscopic confirmations do not correspond to a strong emission
line entering our NB filter. These could be explain by (1) photometric
errors, or (2) an old stellar population, which would decrease the
$z$\arcmin-band flux relative to flux at the NB wavelength
(see Section~\ref{sec:selection}).

We illustrate in Figure~\ref{zNB}, the $\Delta$(\zNB) colors
for spectroscopically confirmed sources. The NB921 spectroscopic
sample spans six magnitudes in luminosity and three magnitudes
in NB921 excess.
Because of sparser sampling, the NB973 spectroscopic sample
covers four magnitudes in luminosity and one magnitude in
NB973 excess.
They show that NB921 (NB973) excesses of 0.2 mag (0.25 mag)
are indeed due to the presence of weak--but real--emission
lines.
%
%
\begin{figure}
  \epsscale{1.2}
  \plotone{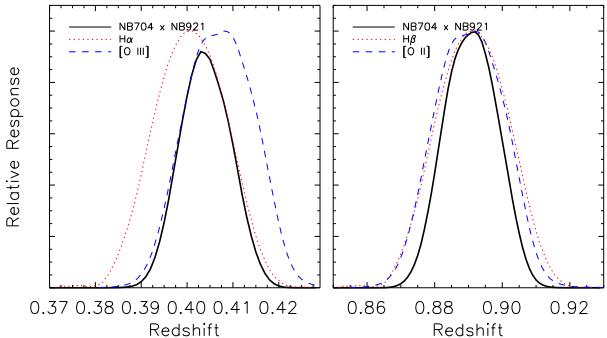}
  \caption{Redshift coverage for dual emitters. Dashed (blue) and dotted
    (red) lines show the NB704 and NB921 bandpasses, respectively. The
    product of the two filters is shown by the solid lines. The left
    (right) figure illustrates the \OIII--\Ha\ (\OII--\Hb) combination
    for $z\approx0.40$ ($z\approx0.89$).}
  \label{dual_em}
\end{figure}
\subsubsection{Dual Line Detection with NB Filter Pairs}
\label{dual_lines}

In addition to spectroscopy, another independent approach showing that
the NB selection technique efficiently identifies line-emitting galaxies, 
is a second NB filter
that targets another emission line at the identical redshift.
Coincidentally, the NB704 filter (centered at 7046\AA\ with a FWHM 
of 100\AA) is able to accomplish this in conjunction with NB921.
Others \citep[e.g.,][]{hippelein03,nagao08,nakajima12} have exploited similar
dual-line instances to confirm emission-line galaxies at a particular
redshift.
When \Ha\ (\Hb) is redshifted into the NB921 filter at $z=0.40$ ($z=0.89$),
\OIII\ (\OII) enters the NB704 bandpass. We illustrate in
Figure~\ref{dual_em} the redshift range probed by these two NB filters
for the \OIII--\Ha\ and \OII--\Hb\ combinations. Cross-matching a
sample of 1421 3$\sigma$-detected NB704 excess emitters
\citepalias{ly07} with the NB921 emitting sample discussed, we found
241 dual NB704-NB921 emitters. These dual emitters are overlaid on
Figure~\ref{zNB} as crosses.
Among the dual emitters, \Ndualz\ were spectroscopically targeted,
and only \Ndualbad\ of them have unidentified redshifts. Among the
remaining 71 sources, 66 of them (93\%)
were \Ha\ emitters.
There are only two cases (3\%) of dual \OII\ and \Hb\ at
$z\approx0.89$. Since these lines are much weaker
\citep[by a factor of 5 to 10; see][]{hicks02} than \OIII\ and
\Ha\ for typical star-forming galaxies at these redshifts, it is
not a surprise that the majority of spectroscopically confirmed
dual emitters are at $z\sim0.4$.
%
%
\subsection{Identifying \OII\ Emitters} 
\label{OIIid}
%
%
\begin{figure*}
  \plottwo{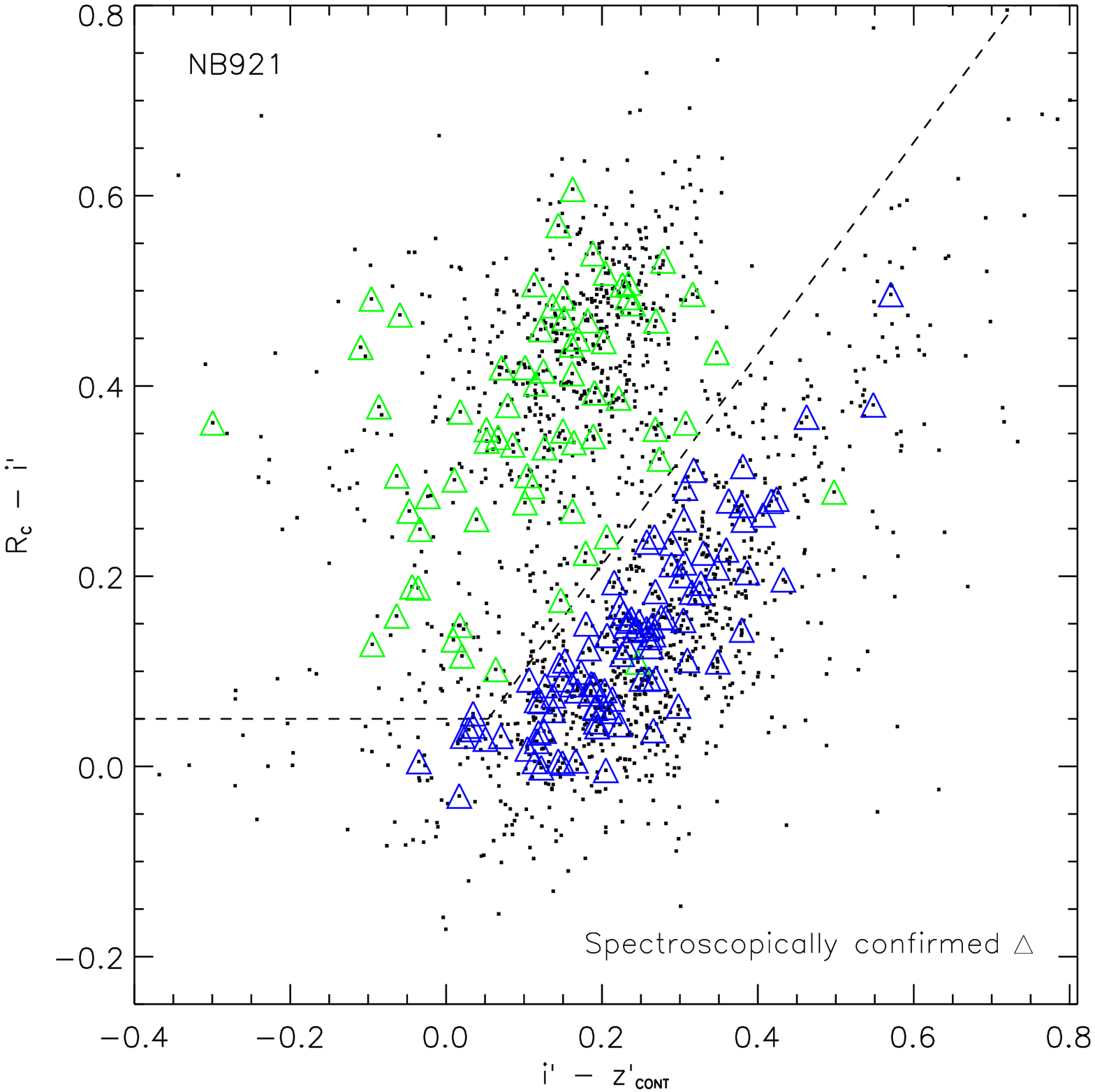}{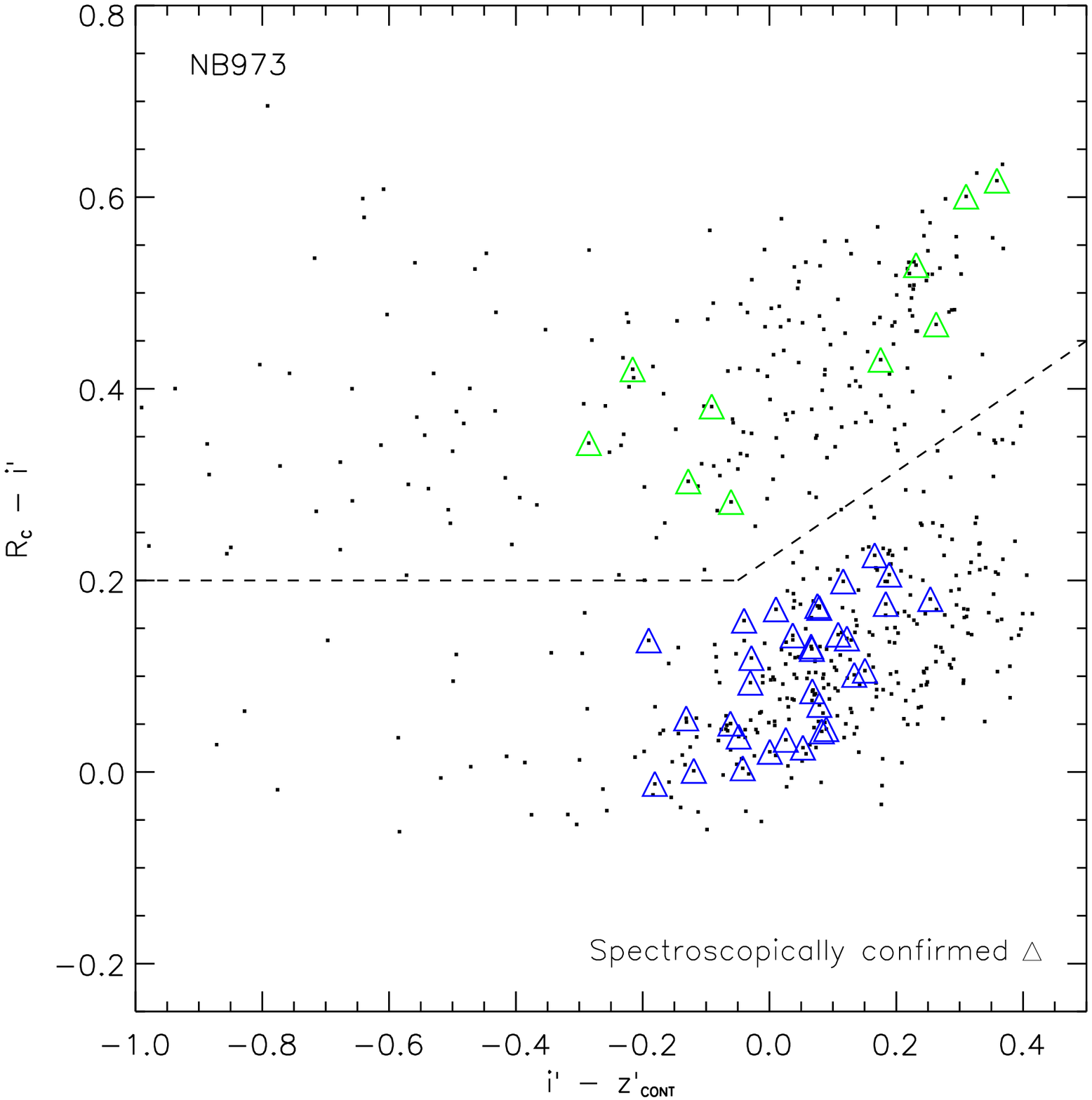}
  \caption{\Rc--$i$\arcmin\ and \iz$_{\rm cont}$ colors for NB921 (left)
    and NB973 (right) emitters shown as black squares.
    \Ha\ emitters, identified in \cite{ly12}, have been excluded from this plot.
    Spectroscopically confirmed galaxies are shown as triangles
    with green and blue colors representing \OIII\ and \OII\ emission
    entering the NB filter, respectively.}
  \label{Riz}
\end{figure*}
Rest-frame optical colors were previously used to distinguish the
three most common redshifted emission lines, \OII, \OIII, and \Ha,
entering into our narrow-band filters \citepalias{ly07}. 
A small sample of spectroscopically confirmed sources is used to
calibrate and optimize the broad-band color selection.
With a much larger spectroscopic sample, we adopt an improved hybrid
approach to select our \OII\ samples for our analysis.
First, spec-$z$'s provide unambiguous classifications of NB excess
emitters. For the NB921 (NB973) sample, \specOIIa\ (\specOIIb)
candidates are definite \OII\ emission-line galaxies.
These spectroscopic confirmations represent $\approx$10\% of the 
final \OII\ samples.
Next, we apply two-color photometry to select \OII\
emitters. Illustrated in Figure~\ref{Riz} are the \Rc--$i$\arcmin\ and
\iz$_{\rm cont}$ colors for our excess emitters, where
$z$\arcmin$_{\rm cont}$ is the ``true'' continuum at $\sim$9200\AA\
after using the NB photometric measurements to remove contamination
of the nebular emission line into the $z$\arcmin:
\begin{equation}
  z\arcmin_{\rm cont} = z\arcmin + 2.5\log{\left(1 + \frac{\rm EW_{obs}}{955\AA}\right)},
\end{equation}
where 955\AA\ is the FWHM of the $z$\arcmin-band filter.
This is especially important since \OIII\ emitters with very
large EWs will have unusually red \iz\ colors, which can be
distorted enough to cross our color selection boundaries.
The median differences in $z$\arcmin\ and $z\arcmin_{\rm cont}$ are
0.09 mag (NB921) and 0.14 mag (NB973), with average differences
of 0.11 mag (NB921) and 0.24 mag (NB973), and can be as large as
1.25 mag (NB921) and 1.62 mag (NB973).
The broad-band color choices were made in \citetalias{ly07}, and were
designed to sample both sides of the Balmer/4000\AA\ break at $z\sim0.9$,
while for higher redshift, probing blueward of the break.
In this figure, we have already excluded \Ha\ emitters with our previous
$B\Rcf i\arcmin$ two-color selection \citep{ly12}.
To select \OII\ NB921 excess emitters with optical colors, we adopt:
\begin{eqnarray}
  \nonumber
  \Rcf-i\arcmin &\leq& 0.05,{\rm~or}\\
  \nonumber
  \Rcf-i\arcmin &\leq& 1.11(i\arcmin-z\arcmin_{\rm cont}) - 0.01.
\end{eqnarray}
These color criteria are similar to those adopted in \citetalias{ly07}.
With a larger spectroscopic sample, we determined that they only miss
1 of 100 (1\%) spectroscopically identified \OII\ NB921 emitters.
Also, only 2 of 70 (3\%) identified \OIII\ emitters meet the 
\OII\ color selection.
For selecting \OII\ NB973 excess emitters, we adopt:
\begin{eqnarray}
  \nonumber
  \Rcf-i\arcmin &\leq& 0.20,{\rm~or}\\
  \nonumber
  \Rcf-i\arcmin &\leq& 0.45(i\arcmin-z\arcmin_{\rm cont}) + 0.22.
\end{eqnarray}
Based on the above broad-band color selection, combined with our
spectroscopic redshifts, we have identified \NOIIa\ \OII\ NB921
at $z\approx\za$ and \NOIIb\ NB973 at $z\approx\zb$.
We will now test these redshift determinations by comparing them
with photometric redshifts.
%
%
\subsection{Photometric Redshifts for NB Excess Emitters}
\label{sec:photoz}
%
%
\begin{figure*}
  \epsscale{1.0}
  \plotone{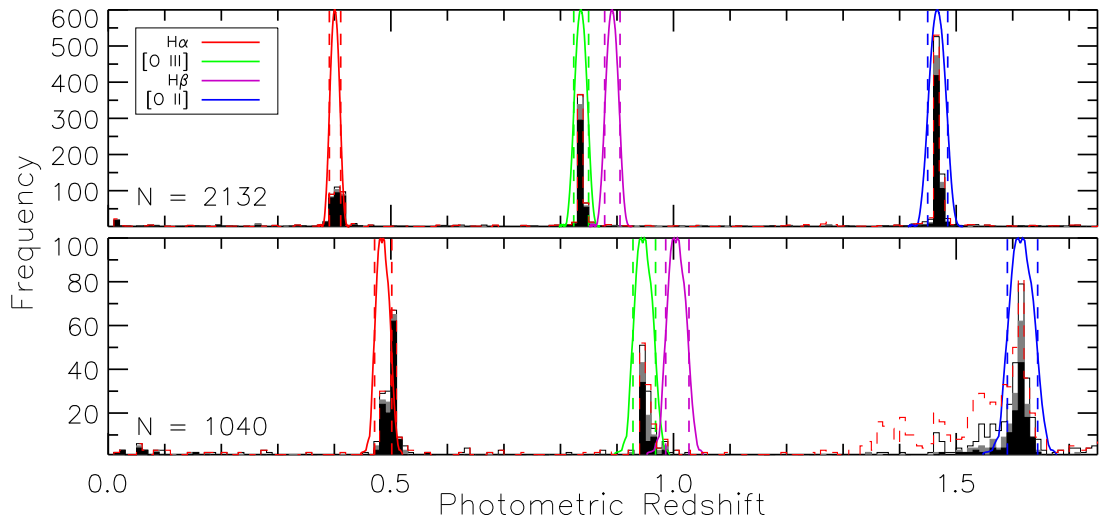}
  \vspace{-3mm}
  \caption{Photometric redshifts for NB921 (top) and NB973
    emitters (bottom) derived from twenty-band photometry.
    Overlaid are the NB921 and NB973 filter profiles at the
    redshifts where \Ha\ (red), \OIII\ (green), \Hb\ (purple),
    and \OII\ (blue) enter the NB filters.  Samples  of NB excesses
    larger than 3$\sigma$, 4$\sigma$, and 5$\sigma$ are shown by
    the solid black line, filled grey, and filled black
    histograms, respectively. The red histograms are intended
    to be compared against the black histograms to denote
    galaxies that have been flagged as interlopers with red \iz\ color
    (see Section~\ref{sec:selection}). Vertical dashed lines show
    the FWHM of the filters.}
    \label{zphot}
\end{figure*}
%
%
\begin{figure*}
  \epsscale{0.55}
  \plotone{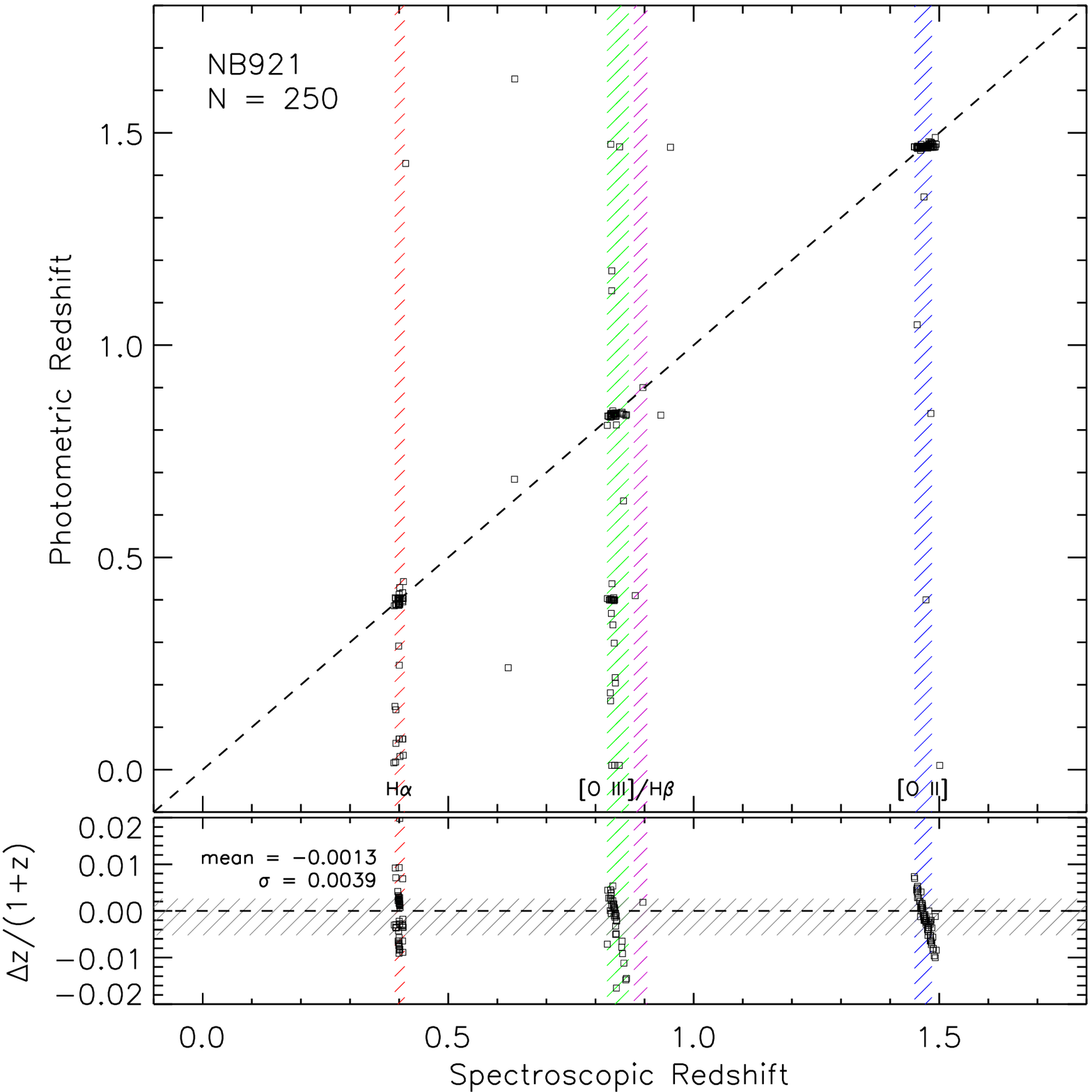}
  \plotone{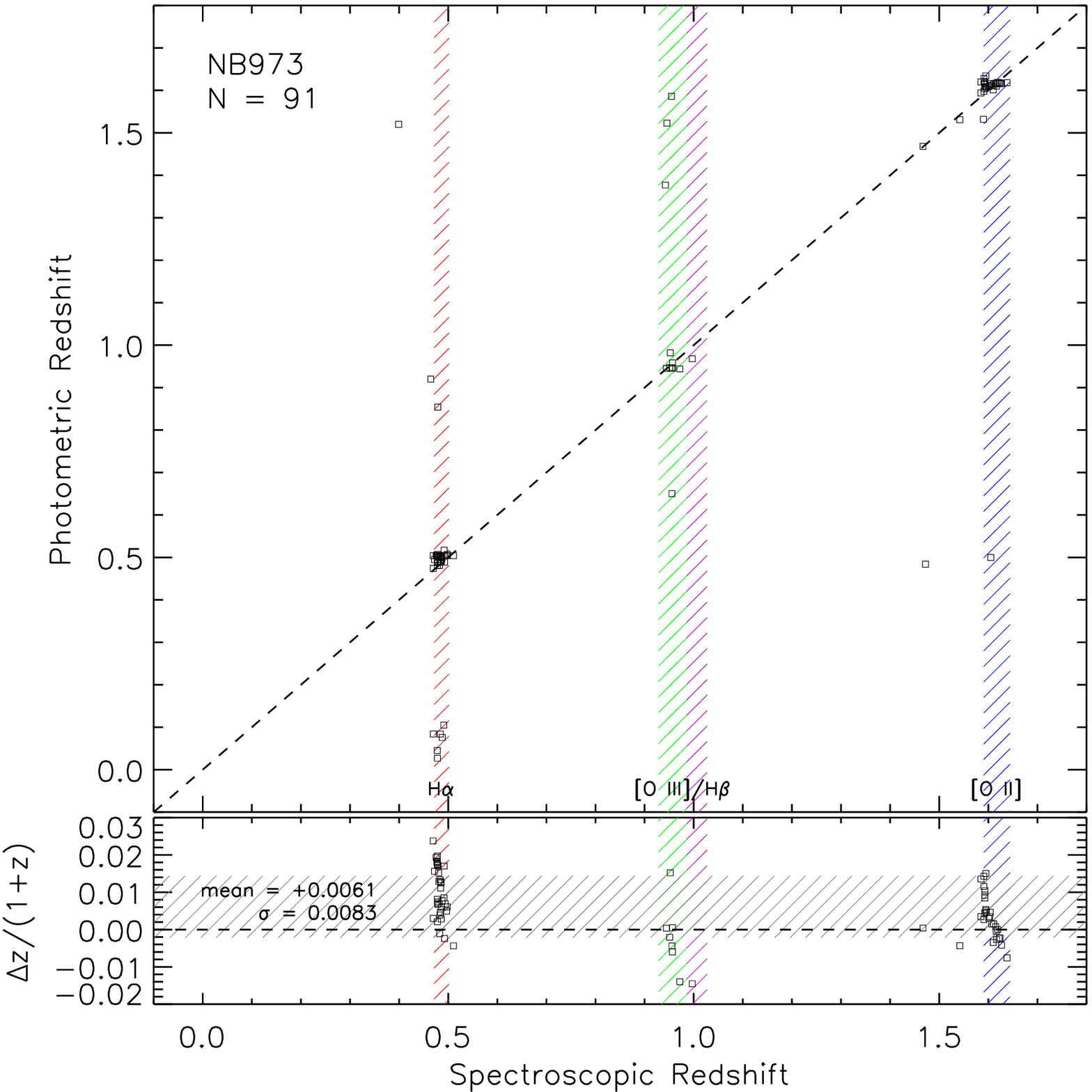}
  \caption{Comparisons between photometric and spectroscopic
    redshifts for NB921 (left) and NB973 (right) excess emitters.
    Shaded regions represent the redshift domain probed by the
    respective NB filter. One-to-one correspondence is shown by
    the dashed line. The bottom panels show the difference
    normalized by 1 + \zspec: (\zphot--\zspec)/(1 + \zspec).}
  \label{zphot_zspec}
\end{figure*}
While spectroscopy is not available for all galaxies,
photo-$z$'s can be determined from their individual SEDs.
In a previous SDF study \citep{ly11a}, we constructed a very large
sample of $\sim$200,000 objects in SDF with
photometric measurements in twenty bands: FUV, NUV,
$UBV\Rcf i\arcmin z\arcmin z_b z_rJHK$, five narrow-band filters
(NB704, NB711, NB816, NB921, and NB973), and two intermediate
bands at $\sim$6500\AA.   
By cross-matching our NB excess emitter catalogs against the
ultra-deep SDF photometric catalog, we have full photometric
measurements for \Nemzaf\ of \Nemaf\ NB921 excess emitters
and \Nemzbf\ of \Nembf\ NB973 excess emitters.
The photo-$z$'s for these sources were derived using the
Easy and Accurate $Z_{\rm phot}$ from Yale \citep[EAZY;][]{brammer08}.
We illustrate these photo-$z$'s in Figure~\ref{zphot}, and 
compare them against our spectroscopic redshifts
in Figure~\ref{zphot_zspec}. There are good agreements
between the photometric and spectroscopic redshifts:
the rms of (\zphot--\zspec)/(1 + \zspec) $\equiv\frac{\Delta z}{1+\zspecf}$
is 0.0039 for NB921 and 0.0083 for NB973.
This is not surprising, because the EAZY fitting
routine includes emission lines, which are needed to
match the observed broad- and narrow-band photometry.
It appears that the photometric redshifts derived for \Ha\ and
\OII\ NB973 excess emitters are systematically off by
$\approx$1\%. The cause is unknown.

Not including NB filter measurements, the photo-$z$ distribution
shows three peaks near the expected location for \Ha, \OIII, and
\OII. However, the photo-$z$ uncertainties are much larger,
for example, \OII\ emitters have \zphot's between 1.4 and 1.8.

\begin{figure}
  \epsscale{1.15}
  \plotone{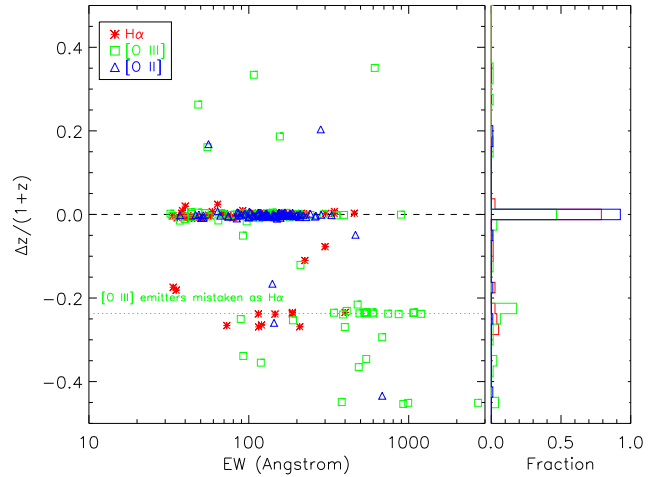}
  \caption{Photometric redshift accuracy, $\frac{\Delta z}{1+\zspecf}$,
    as a function of NB921 observed EW. \Ha, \OIII, and \OII\ emitters
    are denoted as red asterisks, green squares, and blue triangles, 
    respectively. A significant number of the $z=0.84$ \OIII\ emitters have
    incorrect photo-$z$'s due to high \OIII/\Ha\ flux ratios, which
    are not considered in EAZY spectral templates. These
    galaxies, with high \OIII\ EWs, are believed to have low gas
    metallicities.}
  \label{dz}
\end{figure}

We define a \zphot\ outlier as having $\frac{\Delta z}{1+\zspecf} > 0.15$.
There are 56/250 (22.4\%) and 15/91 (16.5\%) failures for the NB921
and NB973 samples, respectively.
As illustrated in Figure~\ref{dz}, the majority of these \zphot\ failures
are associated with galaxies with high observed EWs. For example,
in the NB921 sample, there are almost two dozen spectroscopically
confirmed \OIII\ emitters at $z\approx0.84$ with
EW$_{\rm obs}\gtrsim300$\AA.
All of them have incorrect photo-$z$'s, mostly $\zphotf\sim0.4$.
We determined that the spectral templates
used by EAZY do not allow for extremely strong \OIII\ emission
compared to \Ha\ (i.e., \OIII/\Ha\ flux ratio greater than
unity). As a result, EAZY finds a better fit by assuming
that the large NB921 excess is due to \Ha. As \citetalias{ly07}
have shown, the \OIII/\Ha\ ratio from $z\approx0.4$ dual emitters
were found to be on average, unity.
EAZY will need to be revised to account for the recently observed
fact that star-forming galaxies at high-$z$ have very large
\OIII/\Ha\ ratios, perhaps due to low gas-phase metallicity \citep{atek11}.
A similar problem occurs for the spectroscopically confirmed
\Ha\ emitters where they are placed at $z\sim0.1$, a redshift
which puts \Ha\ in the NB704 filter. However, the NB704 excess
is in fact due to \OIII. This problem occurs when the
observed \Ha+\NII\ EW is above 100\AA.

Other studies have examined photo-$z$'s for emission-line
galaxies and found less catastrophic failures. For example,
\cite{xia11} found very few catastrophic failures between
photometric and grism redshifts. However, it is not clear
what fraction of their sample targeted high EWs, so a
direct comparison is not possible.
We note that there are fortunately very few \OII\ NB921 failures.
Only 3 of 36 (8\%) have incorrect \zphot. The higher success for
the \OII\ selection is because, blueward of 3600\AA, the SED is
simple: no additional emission lines enter our optical bands,
and the Balmer/4000\AA\ break is shifted past 9000\AA.

We note that among the photo-$z$ sample, \POIIzphota\ and \POIIzphotb\
have photo-$z$'s consistent with having \OII\ entering the NB921
and NB973 filters, respectively. These numbers should be taken as lower limits to the
real success of the \OII\ color selection, because many of the
apparently discrepant photometric redshifts were highly uncertain,
based on rather limited multi-band data. If we restrict our
consideration to only the most robustly determined photo-$z$'s, we
would find an even higher success rate of identifying
\OII-emitting galaxies.

%
%


\section{Results}\label{4}

\subsection{\OII\ Equivalent Width Distributions}
\begin{figure}
  \epsscale{1.1}
  \plotone{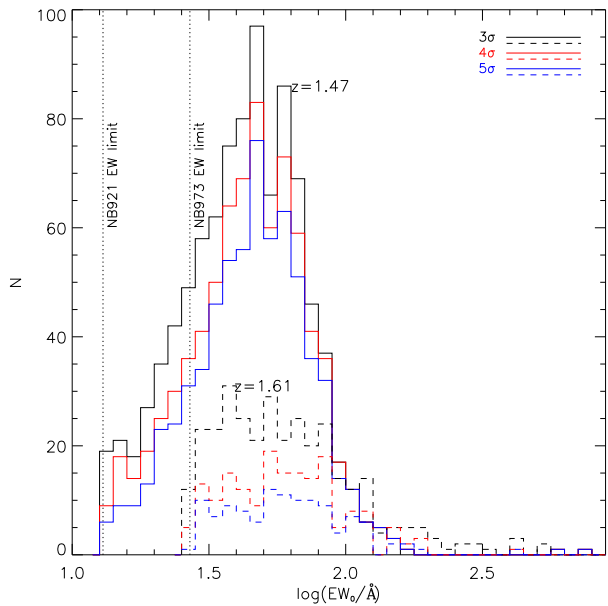}
  \caption{Rest-frame \OII\ EW distributions for our NB921 (solid line)
    and NB973 (dashed line) samples. The dotted lines indicate
    the EW selection limits of our surveys.
    Also shown are the distributions when the samples are limited
    to NB excesses of $>4\sigma$ (red) and $>5\sigma$ (blue), analogous
    to Figure~\ref{zphot}.}
  \label{EWdist}
\end{figure}

Figure~\ref{EWdist} shows the distributions of the rest-frame \OII\ EW
(EW$_0$), for galaxies detected at $\ge3\sigma$ in the NB921 and the NB973 filters.
The distribution for NB921 emitters further emphasizes the excellent
sensitivity our NB921 imaging provides even for galaxies with relatively
weak \OII\ lines, with EW$_0\sim15$\AA.  Although our ability
to detect \OII\ falls off for EW$_0$'s below $\sim$25\AA, the
broad peak of the observed detections around an EW$_0$ of 43\AA\
suggests that our NB921 survey misses relatively few galaxies, since
most of them have EWs larger than our sensitivity limit.
One piece of evidence that supports this is the comparison of
the EW distributions with more conservative NB excesses limits
($4\sigma$ and $5\sigma$; see Figure~\ref{EWdist}).
In these cases, the distributions still have the same shape: peaking
at the same EW with the lower EW slope unchanged. If there was a
significant number of 15--25\AA\ \OII\ emitters detectable at
3--5$\sigma$, we would see a more rapid decline prior to our EW
selection limits.

Recently, \cite{kornei12} also investigated the \OII\ EW of star-forming
galaxies at somewhat lower redshifts of $z=0.7$--1.3. They find a similar
median \OII\ EW$_0$ of 45\AA, and also find a similar 
decline towards lower EWs.
This result provides further independent support for our claim that our
NB921 survey misses few galaxies in the redshift range for \OII, because
that line is nearly always present. 

In addition, another NB survey that examined the \Ha\ EWs at $z=2.2$ finds
that the \Ha\ EW$_0$ distribution peaks at 140\AA,
$\approx$120\AA\ above the survey limit, and intrinsically falls
off rapidly towards lower EWs \citep{lee12}.
%
%

\subsection{Comparison of \OII\ and UV Continuum Luminosities}
For galaxies at any appreciable redshifts ($z>0.5$), the most common
observable used to estimate their current star formation rates (SFRs)
is their UV continuum emission.  This works because in all but the oldest
stellar populations, the far-UV continuum (1500\AA\ rest wavelength)
should be dominated by the hottest, and thus shortest-lived, main-sequence
stars (O and B types).  Since the hottest of these produce most of the 
ionizing photons in a galaxy, it is expected that as the rate of current
star formation increases, so does the UV continuum as well as the
emission line luminosity from \ion{H}{2} regions.  The ``gold standard" 
measure used to count ionizing photons is the \Ha\ recombination line,
but this is redshifted beyond the sensitivity of the best CCD spectrographs
at $z\sim0.5$. To study higher redshifts, secondary emission lines from 
ionized gas have been explored at shorter wavelengths.
The most ``popular back-up'' indicator of SFR is the luminosity of the 
\OII\ doublet.
%
%
\begin{figure*}
  \plotone{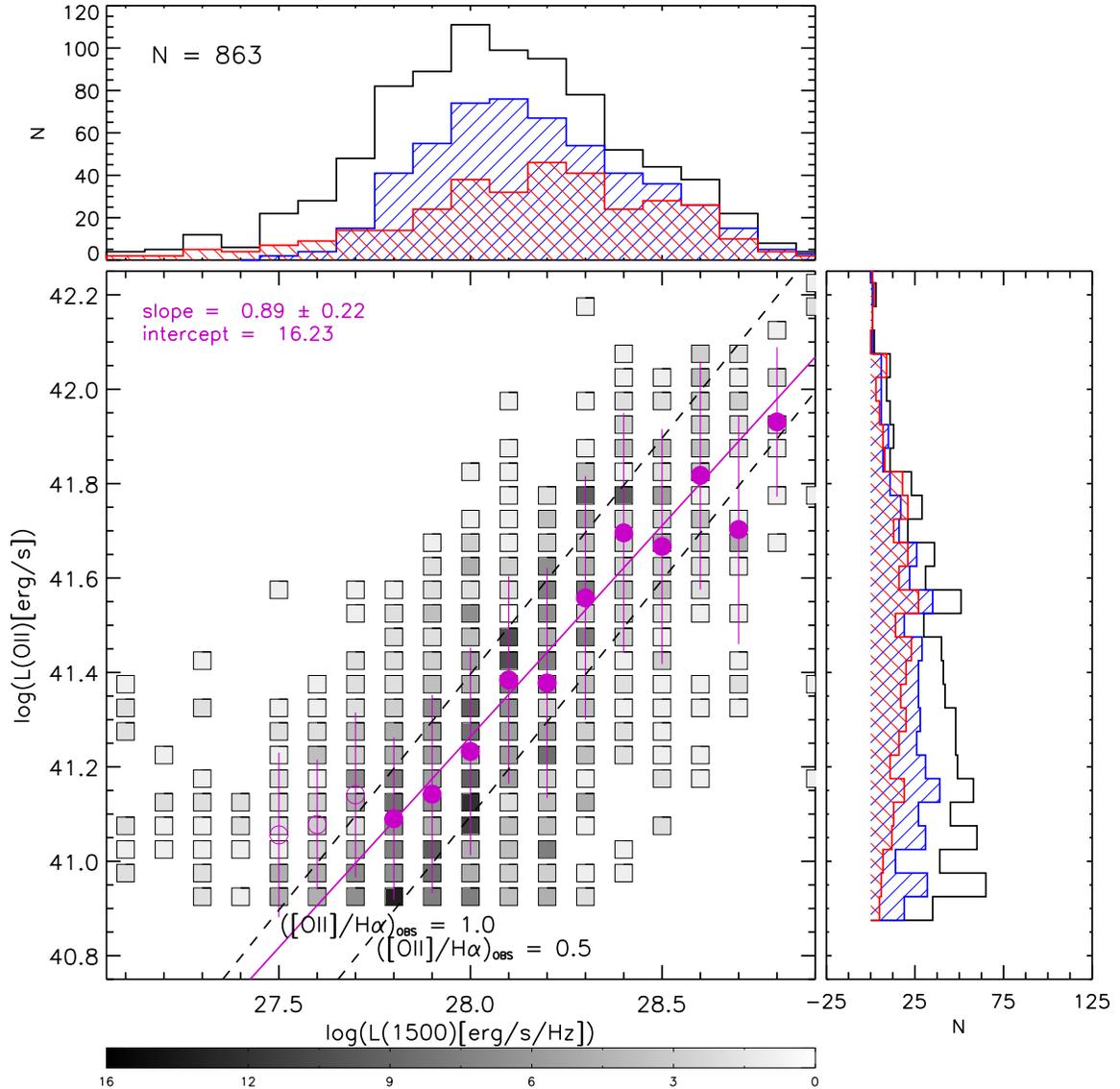}
  \caption{Observed \OII\ and UV luminosities of $z=\za$ \OII\
    emitters. The greyscale represents the density of points
    in each bin. Purple filled circles shown the median at each
    UV luminosity bin with error bars representing 68\% of the
    \OII\ measurements. Overlaid as dashed lines are the cases
    where the observed \OII/\Ha\ flux ratio is 0.5 and 1.0
    following Equation~\ref{eq:OIIUV}. The histograms collapse
    the distributions into one dimension with blue and red colors
    indicating those \OII\ emitters meeting the BX/BM and
    sBzK selections (discussed further in
    Sections~\ref{sec:BXBM}--\ref{sec:sbzk}), respectively.}
  \label{OII_UV}
\end{figure*}

Figure~\ref{OII_UV} shows the {\it observed} luminosities of the \OII\
emission line as a function of the {\it observed} far-UV continuum at
1500\AA.
The solid line shows the least-squares regression
after binning our data in $L$(1500):
$\log{L(\textsc{Oii})} = \SFRint + (\SFRslope) \log{L(1500)}$.
Here, $L$(1500) is determined directly from the $U$-band photometry
\begin{equation}
  \log{\left[\frac{L(1500)}{{\rm erg/s/Hz}}\right]} = \frac{(M_U+48.6)}{-2.5}+40.08,
\end{equation}
where $M_U = U + 2.5\log{(1+z)} - 5\log{(d_L/{\rm 10~pc})}$, and $d_L$
is the luminosity distance, $1.06\times10^4$ Mpc for $z=\za$.
And the \OII\ luminosity is determined from narrow-band and broad-band
measurements:
\begin{equation}
  L(\textsc{Oii}) = \Delta\rNB \left[\frac{f_{\lambda,\rNB}-f_{\lambda,z}}{1-(\Delta{\rNB}/\Delta{z})}\right].
\end{equation}
In fitting the $L$(1500)--$L$(\OII) relation, we have excluded measurements
below $\log{L(1500)}$ = 27.75, since the \OII\ measurements are biased against
galaxies that are fainter than the current \OII\ depth permitted by our
NB imaging.

Both of these observables have been calibrated to predict the SFR
of a galaxy \citep{kennicutt98}, if the quantities can be corrected
for interstellar extinction.
\cite{kennicutt98}'s SFR relation for $L$(\textsc{Oii}) is based on the
calibration for \Ha. In his sample of nearby spiral galaxies, he
estimated $L$(\textsc{Oii}) = 0.45 $L$(\Ha).\footnotemark
\footnotetext[12]{$L$(\textsc{Oii}) and $L$(\Ha) are observed
  (i.e., not de-reddened).}
He recommends an average \Ha\ extinction correction factor of 2.5
\citep{kennicutt92}. Thus, we have
\begin{equation}
  {\rm SFR} = 2.0\times10^{-41} L(\textsc{Oii}) \left[\frac{L(H\alpha)}{L(\textsc{Oii})}\right].
\end{equation}
For a \cite{salpeter} (hereafter Salpeter) initial mass function (IMF),
the far-UV continuum and SFR are also related by a simple linear
proportionality:
SFR = $1.4\times10^{-28} L(1500) \times10^{0.4A(1500)}$,
where $A(1500)=1.37$ mag assuming a \cite{calzetti00} dust curve
with $A$(\Ha) = 1 mag and $\EBV_{\rm star} = 0.44 \EBV_{\rm gas}$.
If we require that both observable diagnostics must yield
the same SFRs, then we would expect a simple linear
relation between observed $L$(\textsc{Oii}) and $L(1500)$:
\begin{equation}
  L(\textsc{Oii}) = 2.5\times10^{13} L(1500)\frac{L(\textsc{Oii})}{L(H\alpha)}.
  \label{eq:OIIUV}
\end{equation}
This linear relation is shown in Figure~\ref{OII_UV} for two values
of the observed \OII/\Ha\ flux ratio, 0.5 and 1.0.
This comparison suggests that, if the \cite{kennicutt98}'s SFR
relations are correct for $L$(1500) and for $L$(\Ha), then our
data indicate that the \OII/\Ha\ flux ratio is typically 0.75
for the majority of our sample. \cite{sobral12} also compared
the \OII\ and \Ha\ flux ratio at the same redshift and found
a typical ratio of 0.4--0.8. For a subset of our sample with high
UV luminosities, the \OII/\Ha\ ratio decreases towards 0.5.
This is consistent with the same decline in \OII/\Ha\ flux ratio
with blue stellar luminosity found by \cite{jansen01} in local
galaxies. Illustrated in Figure~\ref{OII_UV}, the scatter of the
\OII\ luminosity relative to the UV luminosity suggests that
\OII\ can be used as a SFR indicator with a reliability of $\sigma=0.23$ dex.

\subsection{Average Spectral Energy Distributions of \OII-Emitting Galaxies}
\label{SED_avg}
We have combined our multi-wavelength photometry  from hundreds of \OII-emitting
galaxies to construct representative broad-band SEDs.
We limit our results to the NB921 sample, since the
results are better constrained with the larger and deeper sample.
Our filters, $U$ through IRAC 1 and 2 \citep{fazio04}, span rest-frame
1500\AA--1.8\mm, more than ten times the spectral coverage that was
possible with the Suprime-Cam data alone, that was used by \citetalias{ly07}.

\begin{figure*}
  \plottwo{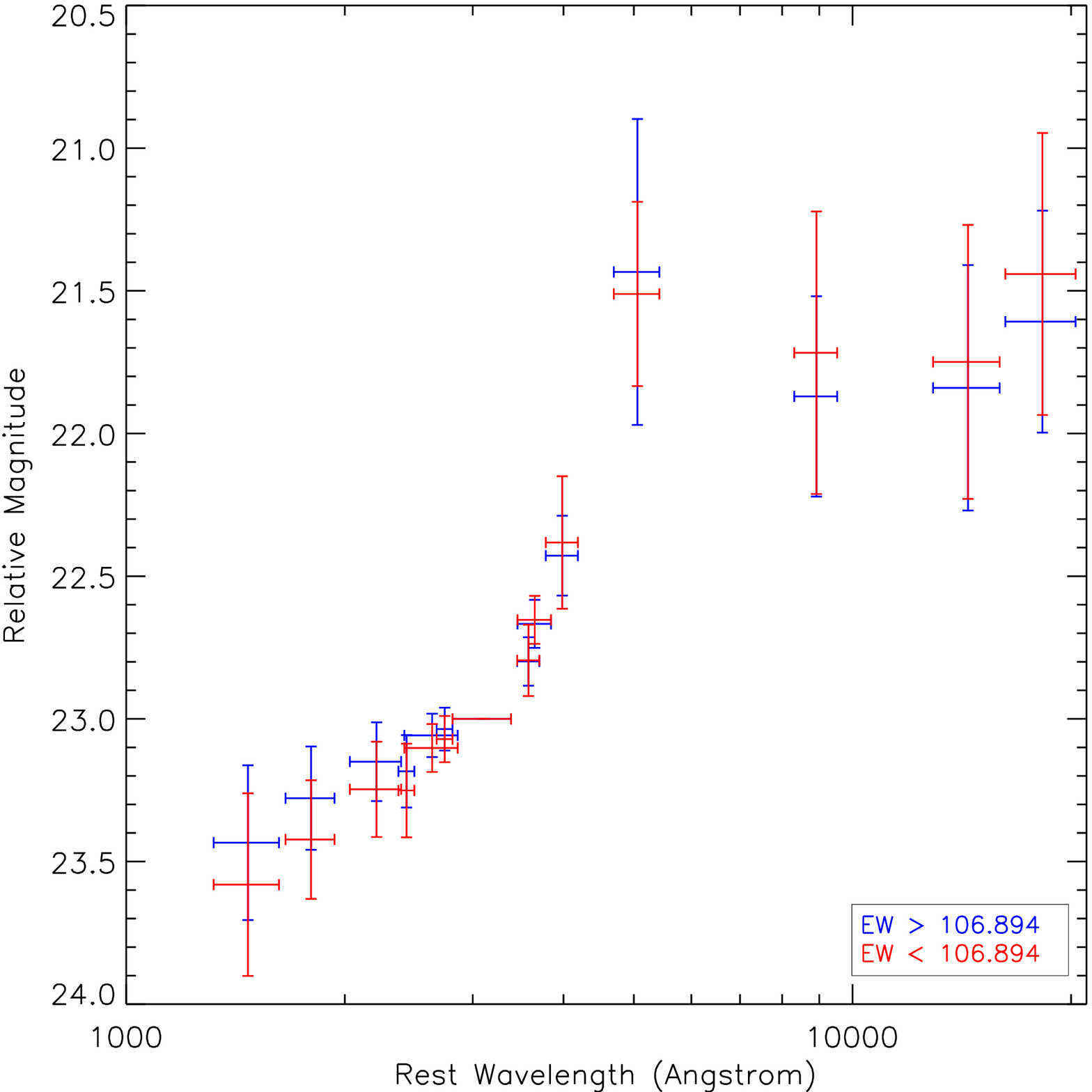}{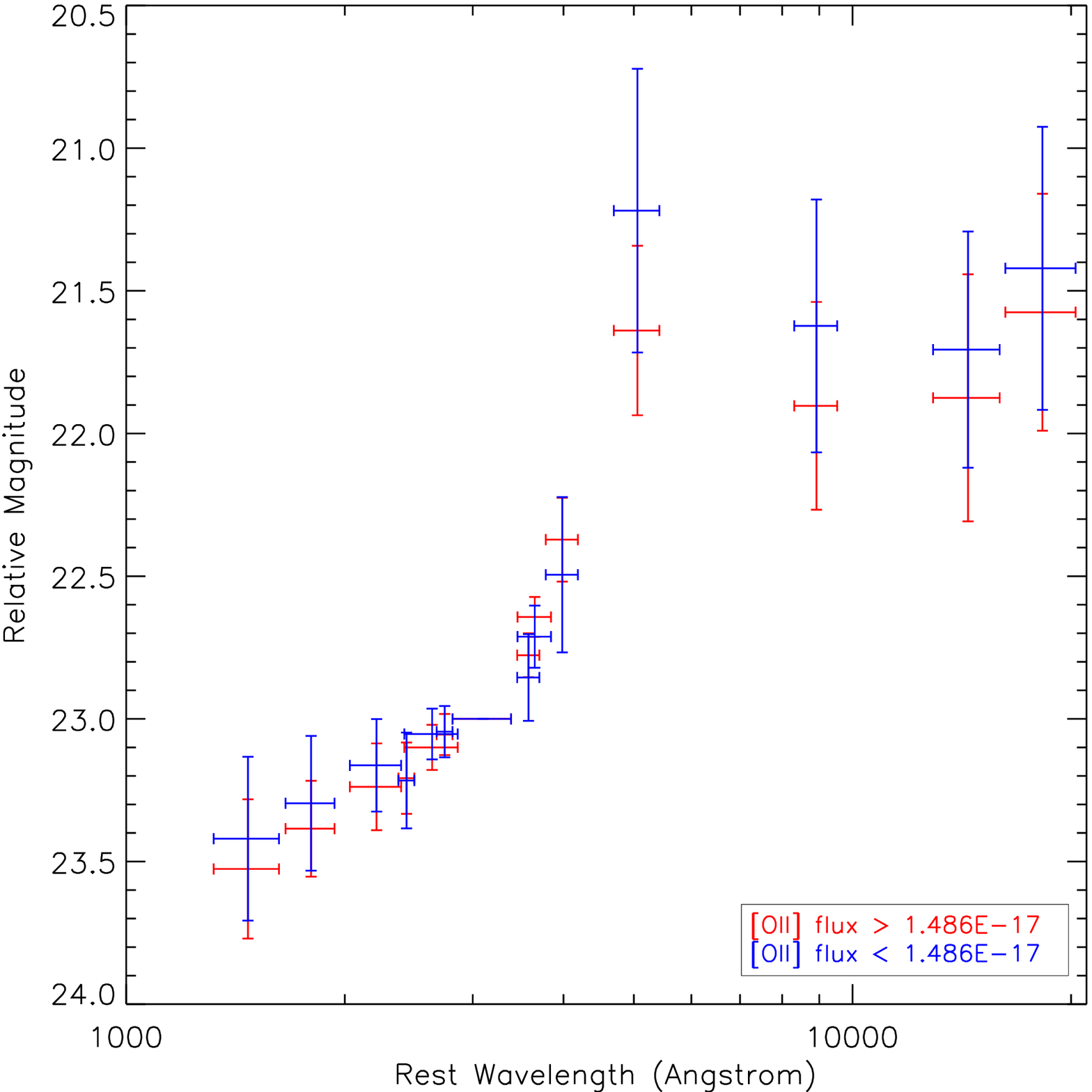}
  \caption{Broad-band SEDs for NB921 \OII\ emitters constructed
    by binning the sample in EW (left) and \OII\ emission-line
    flux (right). Each SED is constructed from roughly 350
    galaxies where the sub-samples are constructed above and below
    the median EW and \OII\ flux. Error bars represent 68\% of the
    sub-samples, and blue and red colors denote the samples
    with bluer and redder rest-frame UV colors.}
  \label{SED}
\end{figure*}

In Figure~\ref{SED} we show the median SEDs of
$z=\za$ \OII\ emitters, dividing the sample between
high- and low-equivalent width of \OII\ emission
(greater or less than 106\AA) and between low and high
\OII\ flux (above and below $1.49\times10^{-17}$ erg s$^{-1}$ cm$^{-2}$).
For comparative purposes, all SEDs are normalized to 23.0
mag in the $i$\arcmin\ band. The error bars show the range
encompassing 68\% of the data in each waveband.

The average $U$-to-$i$\arcmin\ photometry (corresponding
roughly to rest wavelengths between 1500 and 3300\AA)
can be approximated by a power law ($F_\nu \propto \nu^{\alpha}$)
with $\alpha = -0.64$, except for a marginal but systematic dip
in the intermediate 6000 \AA\ filter, which might possibly be
attributed to the broad 2200 \AA\ peak in the interstellar
reddening law.
On the red side of the $i$\arcmin\ band, the SEDs rise much more
rapidly, due to the Balmer and \ion{Ca}{2} H and K breaks with
an average $i\arcmin-K$ color of 1.2 mag.
Redward of the jumps, the rest-frame optical continuum is relatively
flat. This is supported by the IRAC [3.6] and [4.5] photometry.
We note also that at this redshift of $z=1.47$,
\OIII\ and \Hb\ enter the $J$-band, and thus can alter these
colors if these emission-line EWs are extremely
high, as seen in other surveys \citep[e.g.,][]{atek11}.
Indeed, if we compare the average $J$ flux of the high-EW
galaxies to a typical model
SED going through the surrounding filters, it appears to be
a few tenths of a magnitude brighter.  This is attributable to
the contamination from the \OIII\ and \Hb\ lines.

As found by \citetalias{ly07}, the average SED of the high-EW sub-sample is
systematically bluer than of the moderate-EW subset: the observed
$<U-i\arcmin>$ and $<i\arcmin-K>$ colors are $0.43\pm0.27$ and
$1.13\pm0.35$ and $0.58\pm0.32$ and $1.28\pm0.50$ for the high- and
low-EW sub-samples, respectively.
This is reasonable, since the former galaxies, with specific SFRs
several times higher, should have a larger fraction of younger, blue stars.
We also find that galaxies with higher \OII\ emission-line fluxes are
found to be redder than those with weaker \OII\ flux. We argue that
this is due to a selection effect that more massive galaxies tend
to be redder will have higher SFRs and thus higher \OII\ fluxes.
We next quantify these trends with fits of model stellar
populations to the SEDs.
%
%

\subsection{Stellar Populations from SED Modeling}
We modeled the fifteen-band SEDs of our \OII\ emitters to
derive stellar masses, SFRs, dust reddening ($A_V$'s), and
stellar ages. At $z\sim1.5$, the data span rest-frame
600\AA--8800\AA.
The spectral synthesis models are discussed further in \cite{ly11a}.
In brief, we use the IDL-based code, Fitting and Assessment of Synthetic
Templates \citep[FAST;][]{kriek09}, with exponentially declining
$\tau$ star-formation history models from \cite{bc03}.
We adopt a \citetalias{salpeter} IMF, and dust
attenuation follows the \cite{calzetti00} reddening formalism. A
\cite{chabrier03} IMF was also considered; however, the resulting
$\chi^2$ fits (ages, $\tau$'s, and $A_{V}$'s) were no different
from those determined using a \citetalias{salpeter} IMF. The only
difference was a lower stellar mass and SFR by 0.25 dex (a factor of
1.8), which is expected due to the smaller proportion of low-mass stars.
Therefore, we adopt the \citetalias{salpeter} IMF for the remainder of this paper.
To correct for neutral hydrogen intergalactic medium absorption, we
follow the prescriptions described in \cite{madau95}.

The grid of models consists of $\log{(\tau/{\rm yr})} = $ 7.0, 8.0, 9.0,
and 10.0. These $\tau$ values were selected to include a bursty,
intermediate, and roughly constant star-formation history. In addition
to the star-formation history, the grid of models spans a range of
$\log{({\rm age/yr})}$, between 6.0 and up to the age of the universe
at a given redshift (in increments of 0.1 dex), and dust extinction with
$A_V=0.0$--3.0 mag with 0.1 mag increments. We only consider solar
metallicity models, since distinguishing different metallicities is
nearly impossible with broad-band SED fitting \citep[see e.g.,][]{pforr12}.

In these fits, the redshifts are accurately fixed, either from the
spectroscopic values when available, or from the presence of
\OII\ in the NB bandpass ($z=1.467$ or $z=1.617$).
Since these spectral synthesis models do not include nebular emission lines,
we correct the $z$\arcmin-band measured fluxes to remove the 
\OII\ emission contribution to these filters. We note that by having
the photometry as is, the stellar masses were higher by 0.2 dex, the
stellar ages were older by 0.1 dex, and the SFRs were lower by 0.3 dex.

\begin{figure*}
  \epsscale{1.15}
  \plottwo{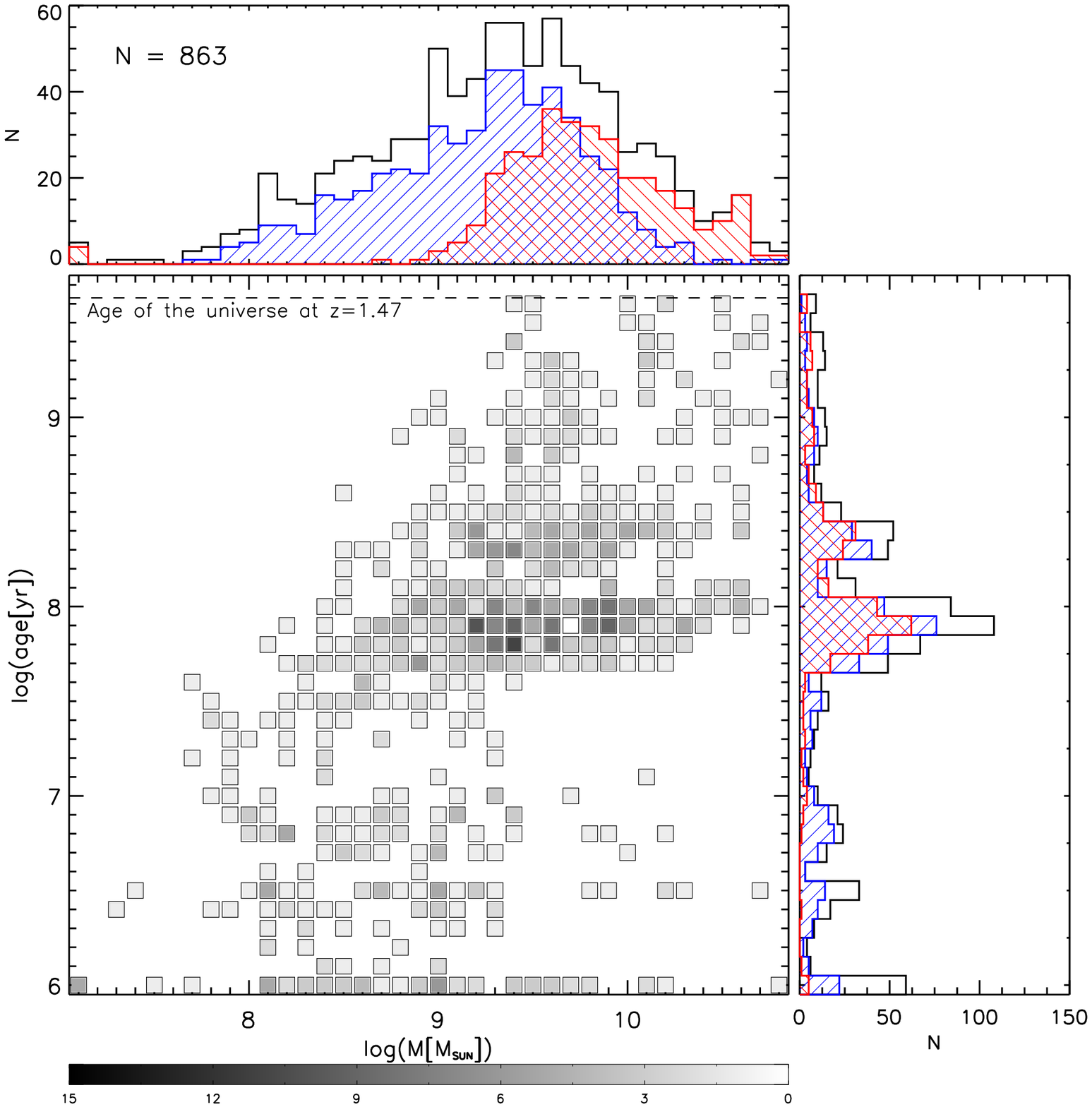}{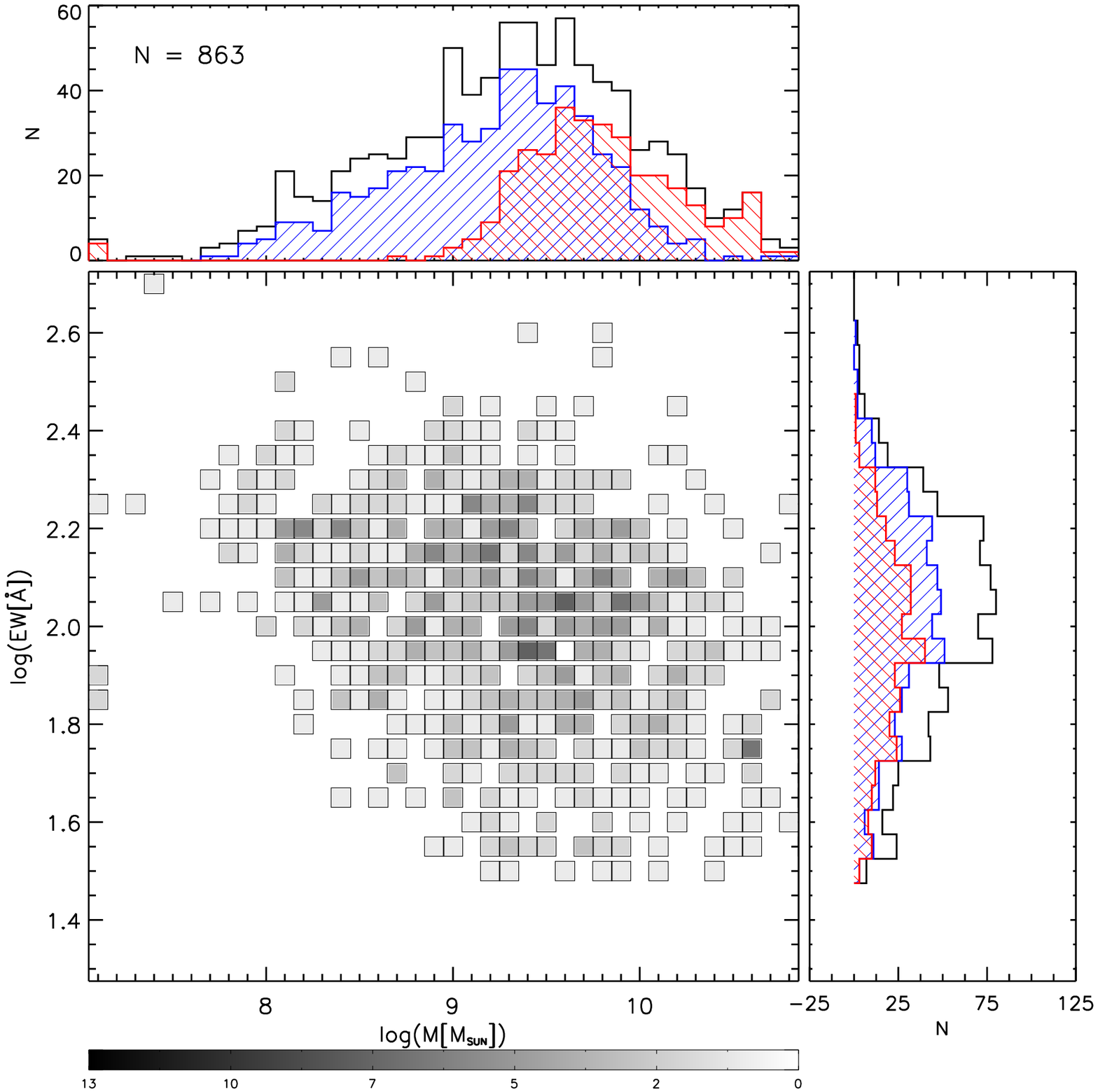}
  \caption{Stellar ages, stellar masses, and \OII\ observed EWs for the
    $z=\za$ \OII\ NB921 emitters. These quantities were derived from SED
    modeled fits that assumed a \citetalias{salpeter} IMF, an exponentially
    declining star-formation history, and \cite{calzetti00} internal
    dust reddening. The
    histograms collapse the distributions into one dimension with blue 
    and red colors indicating those \OII\ emitters meeting the BX/BM and
    sBzK selections, respectively.}
  \label{stellar_pop}
\end{figure*}

In Figure~\ref{stellar_pop}, we illustrate the stellar masses and
ages for 863 $z\approx\za$ \OII\ emitters. We find that these \OII\
emitters span 2 dex in stellar age with a median of $\sim$100 Myr
and 3 dex in stellar mass with a median of $3\times10^9$ \Msun.
In modeling the broad-band SED, there is a known degeneracy between
the fitted stellar age and dust reddening in describing the overall
spectral slope, particularly in the UV. In particular, one can make
galaxies appear younger but with significant dust and vice versa.
We caution that fits with very young stellar populations
($\lesssim$10 Myr) may not be reliable because they also give very
large extinction ($A_V = 1$--3 mag).

We find that the galaxies with observed \OII\ EW above 106\AA\ have on
average a factor of two less stellar mass than the low-EW galaxies.
In addition, galaxies with \OII\ fluxes above $1.49\times10^{-17}$ erg
s$^{-1}$ cm$^{-2}$ are more massive by a factor of 5. These results
are consistent with the average SED results discussed in
Section~\ref{SED_avg} since more (less) massive galaxies are typically
expected to have redder (bluer) colors.
%
%

\subsection{Testing Color-Selection Techniques at $z>1$}
The two techniques commonly used to select star-forming galaxies
at redshifts of 1--3 are the ``BX/BM'' \citep{adelberger04} and ``sBzK''
\citep{franx03,daddi04} methods. The former approach, analogous
to the Lyman break method \citep{steidel92}, finds young galaxies
with strong star formation and moderately red $\Un-G$ color due to
the Lyman continuum break beginning to enter into the $\Un$
filter.
The latter method is intended to identify $z\gtrsim1.4$ galaxies when the
Balmer/4000\AA\ break occurs between $z$\arcmin\ and $K$.
Because of the need for $K$-band data, the BzK method is
generally biased towards more massive galaxies.
These methods were developed to help ``fill'' the historically known
``redshift desert'' where wide-field deep UV imaging to identify
$z\sim1.5$ Lyman break galaxies was not available until more
recently \citep{ly09,hathi10,oesch10,swift11,haberzettl12}.
These color selections and their respective samples identified
for the SDF are discussed in \cite{ly11a}, where they illustrated
that a wide range of the stellar populations of galaxies at $z\gtrsim1$
is probed by combining these two complementary methods.

With deep multi-wavelength coverage for the SDF, we examine which of
our \OII\ emitters are classified as
UV-selected ``BX/BM'' galaxies and/or NIR-selected BzK galaxies.

Since we argued above that {\it most} galaxies at $z=\za$
have detectable \OII\ emission lines, we can use our large
NB samples to test the completeness of the two main
broad-band color search methods available at
the same redshift.  We find that both suffer from some incompleteness,
which is mostly eliminated by combining the two methods.
%
%
\subsubsection{The BX/BM Sample}\label{sec:BXBM}
%
%
\begin{figure*}
  \epsscale{1.1}
  \plottwo{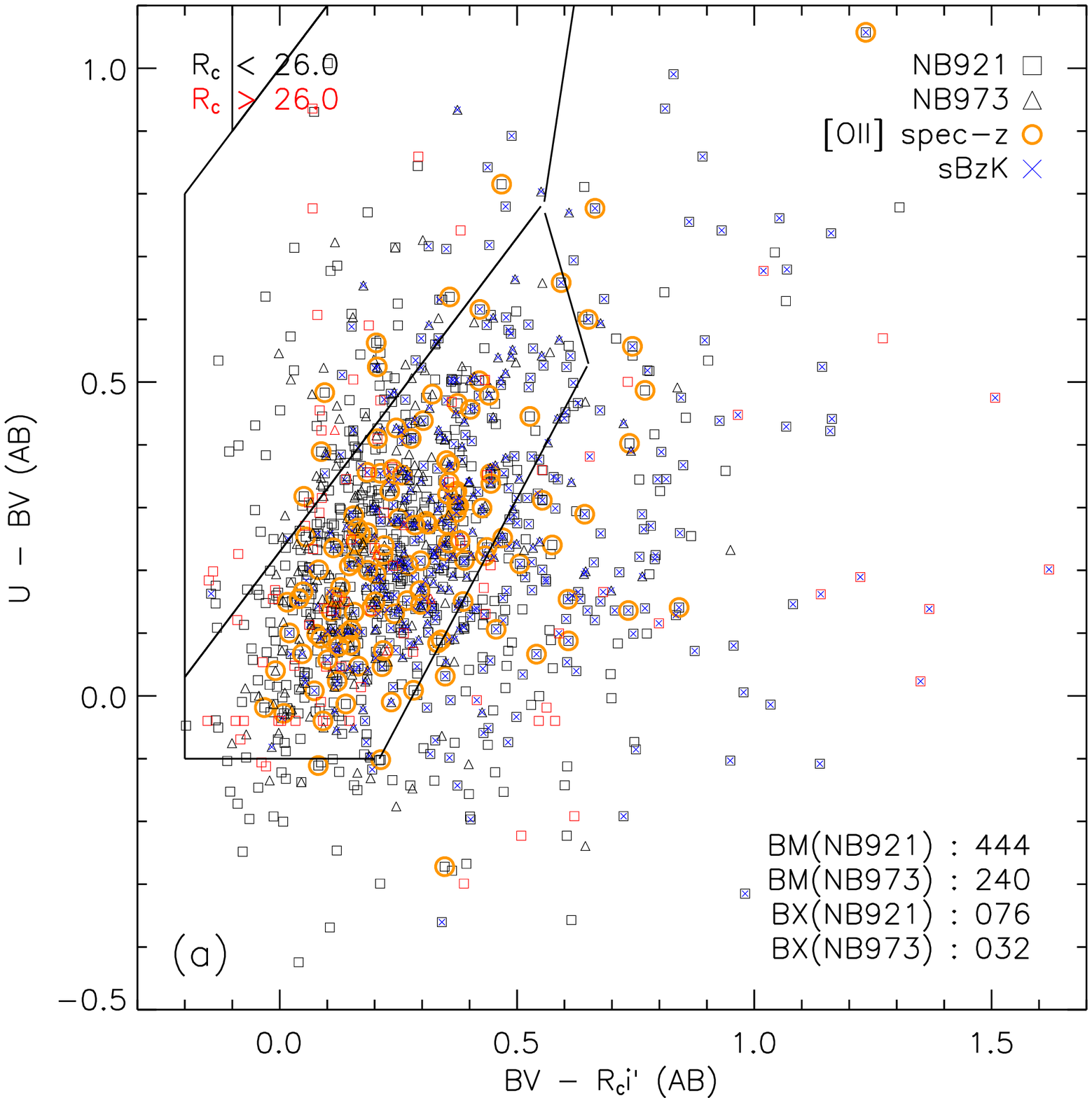}{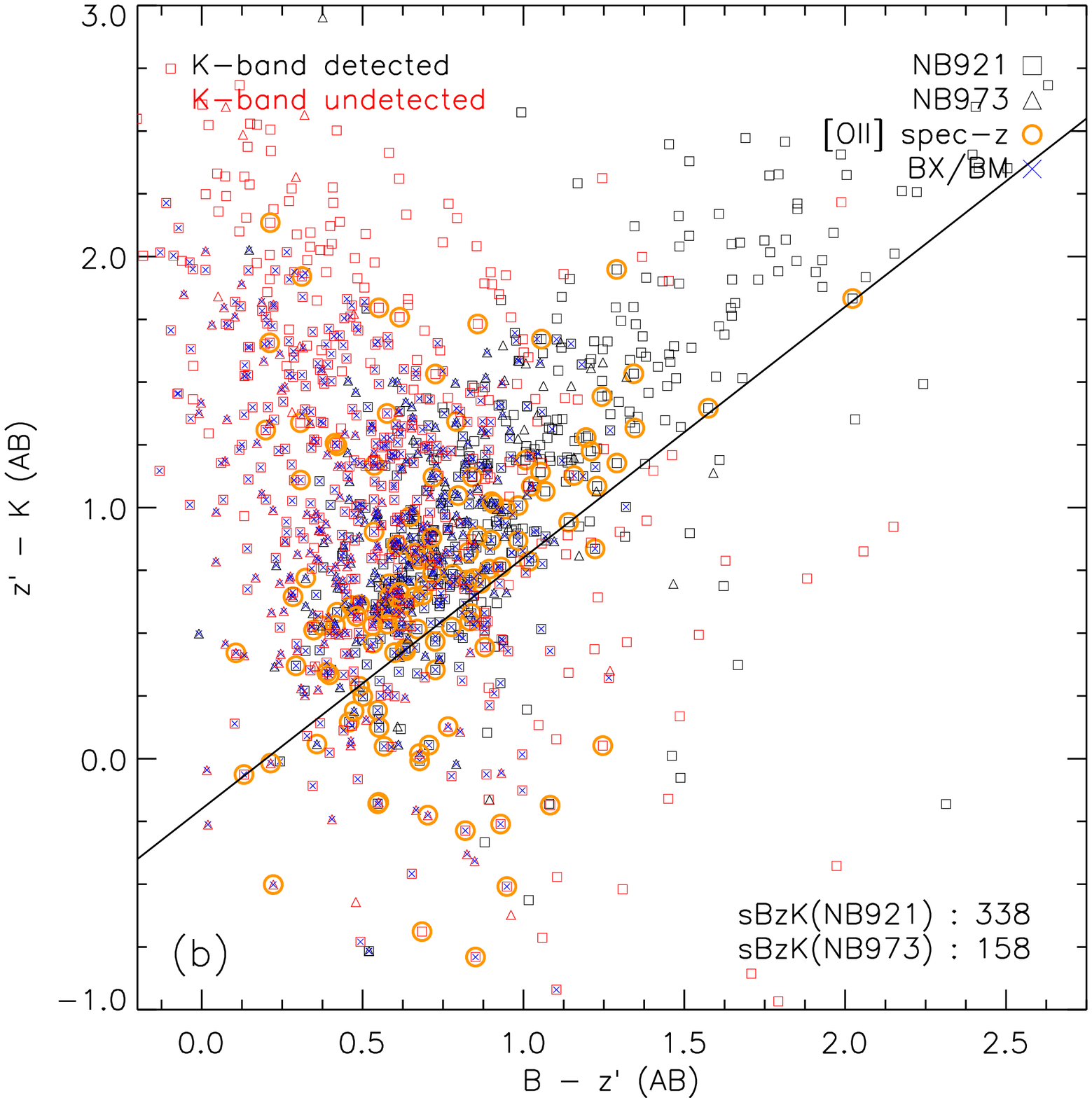}
  \caption{({\it a}) \UBV\ and \BVRi\ colors for \OII\ emitters. Red symbols
    denote sources fainter than $\Rcf=26.0$. BM, BX, and $U$-dropout
    galaxies are selected by the three regions defined by the solid lines
    (from bottom to top, respectively). Most galaxies are classified
    as BM, which is expected for $z=\za$\ and \zb. ({\it b})
    $B$--$z$\arcmin\ and $z$\arcmin--$K$ colors for \OII\ emitters. Red
    symbols identify sources that are undetected in the $K$-band at the
    3$\sigma$ level (upper limits on $z$\arcmin--$K$). Star-forming BzK
    galaxies are classified as those lying above the diagonal line.
    Both panels show blue crosses to denote the galaxy population
    selected by the other method. The NB921 and NB973 \OII\ samples are
    distinguished by squares and triangles, respectively. Spectroscopically
    confirmed \OII\ emitters are shown as orange circles.}
  \label{UGR_BzK}
\end{figure*}

Figure~\ref{UGR_BzK}(a) shows the $\Un-G$ and $G-\Rs$ colors for both of
\OII\ emitting samples.
Since the SDF does not have these exact filters, we derive these colors
by using a combination of the $UBV\Rcf i\arcmin$ filters that are measured.
The color transformation was determined by using stellar population synthesis
models representative of UV luminous galaxies \citep{ly11a}, and consists of:
\begin{eqnarray}
\label{eqnBV}
\Un &=& U,\\
G   &=& BV \equiv -2.5\log{\left[\frac{x_1f_B + (1-x_1)f_V}{3630~\mu{\rm Jy}}\right]},\textrm{~and}\\
\label{eqnRi}
\Rs &=& \Rcf i\arcmin \equiv -2.5\log{\left[\frac{x_2f_R + (1-x_2)f_{i\arcmin}}{3630~\mu\textrm{Jy}}\right]}.
\end{eqnarray}
Here, $f_X$ is the flux density per unit frequency (erg s$^{-1}$ cm$^{-2}$
Hz$^{-1}$) in band ``\textit{X}'', and $x_1=0.314$ and $x_2=0.207$, as
discussed in \cite{ly11a}.

Limiting our consideration to $\Rcf \leq 25.5$, the typical depth adopted by
past and current BX/BM surveys, we find for the full $z=\za$ \OII\ sample that
47\% are BX or BM. Likewise, 82\% of the $z=\zb$ \OII\ emitters are either
BX or BM. We summarize these statistics, as well as those for $\Rcf \leq 26.0$,
in Table~\ref{tab:BXBM_BzK}.

\begin{deluxetable*}{cccccccccc}
  \tabletypesize{\scriptsize}
  \tablewidth{0pc}
  \tablecaption{Summary of BX/BM and \sbzk\ Results for \OII\ Emitters}
  \tablehead{
    \colhead{Filter}&
    \colhead{$N$}&
    \colhead{$N_{\rm phot}$}&
    \colhead{$N(R\leq25.5)$}&
    \colhead{$N(R\leq26.0)$}&
    \colhead{UV(R$\leq$25.5)}&
    \colhead{UV(R$\leq$26.0)}&
    \colhead{$N$(K-det)}&
    \colhead{$N$(sBzK)}&
    \colhead{$N$(Either)}\\
    \colhead{(1)}&\colhead{(2)}&\colhead{(3)}&\colhead{(4)}&\colhead{(5)}&\colhead{(6)}&
    \colhead{(7)}&\colhead{(8)}&\colhead{(9)}&\colhead{(10)}}
  \startdata
  NB921  & 933 & 863 92.5\% & 585 67.8\% & 741 85.9\% & 406 47.0\% & 520 60.3\% &     \ldots & \ldots     & \ldots      \\
  \ldots &(814)&(750 92.1\%)&(505 67.3\%)&(641 85.5\%)&(348 46.4\%)&(446 59.5\%)&(402 53.6\%)&(338 45.1\%)& (596 79.5\%)\\
  NB973  & 328 & 313 95.4\% & 286 91.4\% & 307 98.1\% & 255 81.5\% & 272 86.9\% &     \ldots &     \ldots & \ldots      \\
  \ldots &(278)&(264 95.0\%)&(238 90.2\%)&(258 97.7\%)&(210 79.5\%)&(226 85.6\%)&(174 65.9\%)&(158 59.8\%)& (248 93.9\%)\\
  \vspace{-3mm}
  \enddata
  \label{tab:BXBM_BzK}
  \tablecomments{Values reported in parentheses refer those sources that fall in
    the sensitive $K$-band region (``K-det''). Except for those reported in Col.
    (3), all percentages are relative to $N_{\rm phot}$. In Cols.(6)--(7),
    ``UV'' refers to the combination of ``BX'' and ``BM'' selections.}
\end{deluxetable*}

%
%
\subsubsection{The \sbzk\ Sample}\label{sec:sbzk}
The $B-z\arcmin$ and $z\arcmin-K$ colors for the \OII\ emitters are illustrated
in Figure~\ref{UGR_BzK}(b). Since the SDF is not fully covered (80\%) with
UKIRT/$K$-band observations to a 3$\sigma$ sensitive depth of 23.9 mag, we
limit the sample of \OII\ emitters to those that are in the sensitive region.
Among the 863 $z=1.47$ (313 $z=1.61$) \OII\ emitters, \NOIIaK\ (\NOIIbK)
sources are located within the sensitive region.
Requiring a 3$\sigma$ detection in $K$, the samples are further limited to
\NOIIaKdet\ (NB921) and \NOIIbKdet\ (NB973). \cite{daddi04} defines
star-forming BzK (sBzK) as galaxies with $z\arcmin-K \ge (B-z) - 0.2$.
This yields \sbzkap\ of \OII\ NB921 and \sbzkbp\ of \OII\ NB973 emitters as
sBzK galaxies. None of these \OII\ emitters have colors that would classify them as
passive BzK. Thus both color methods, especially BzK, work very well to find
emission-line galaxies.

We emphasize that 80 of 576 (14\%) $K$-band detected \OII\ emitters have
blue $z\arcmin-K$ colors preventing them from being selected as sBzKs.
As illustrated in  Figure~\ref{UGR_BzK}(b), a large fraction of these has
been spectroscopically confirmed to be at $z=\za$ or $z=\zb$.
\cite{lee12} have also shown that for a sample of $z=2.2$ \Ha\ emitters,
8\% are also missed because of their blue $z\arcmin-K$ colors. They argued
that the sBzK selection fails to identify galaxies with relatively younger
stellar population due to little or no Balmer break.
\cite{barger08} also examined the BzK selection with a large
spectroscopic sample and find that the sBzK selection is effective
at identifying high redshift galaxies, but may suffer from
significant (33\%) contamination from $z<1.4$ interlopers.

%
%
\subsubsection{Stellar Population Characteristics of Color-Selected Galaxies}
\label{sec:4.5.3}
To illustrate the diversity of the \OII\ sample and the selection
biases associated with the BX/BM and sBzK techniques, we
plot their distributions of stellar masses, ages, and \OII\ EW
in Figure~\ref{stellar_pop}. It is apparent that the sBzK selection
(shown in red) is well designed to identify almost all of the
massive \OII\ emitters, but becomes seriously incomplete
below $3\times10^9$ \Msun. This incompleteness is caused by the
current depth in the $K$-band.
By contrast, the BX/BM selection (shown in blue), succeeds at identifying
most of the low-mass galaxies, but misses about half of the massive
population. In addition, we present the first determination of \OII\ EW
distributions for large representative samples of BX/BM and sBzK galaxies.
The BX/BM sample is skewed towards higher observed EWs
(average of 105\AA) compared to sBzKs (average of 91\AA).

%
%
\begin{figure*}
  \epsscale{1.1}
  \plottwo{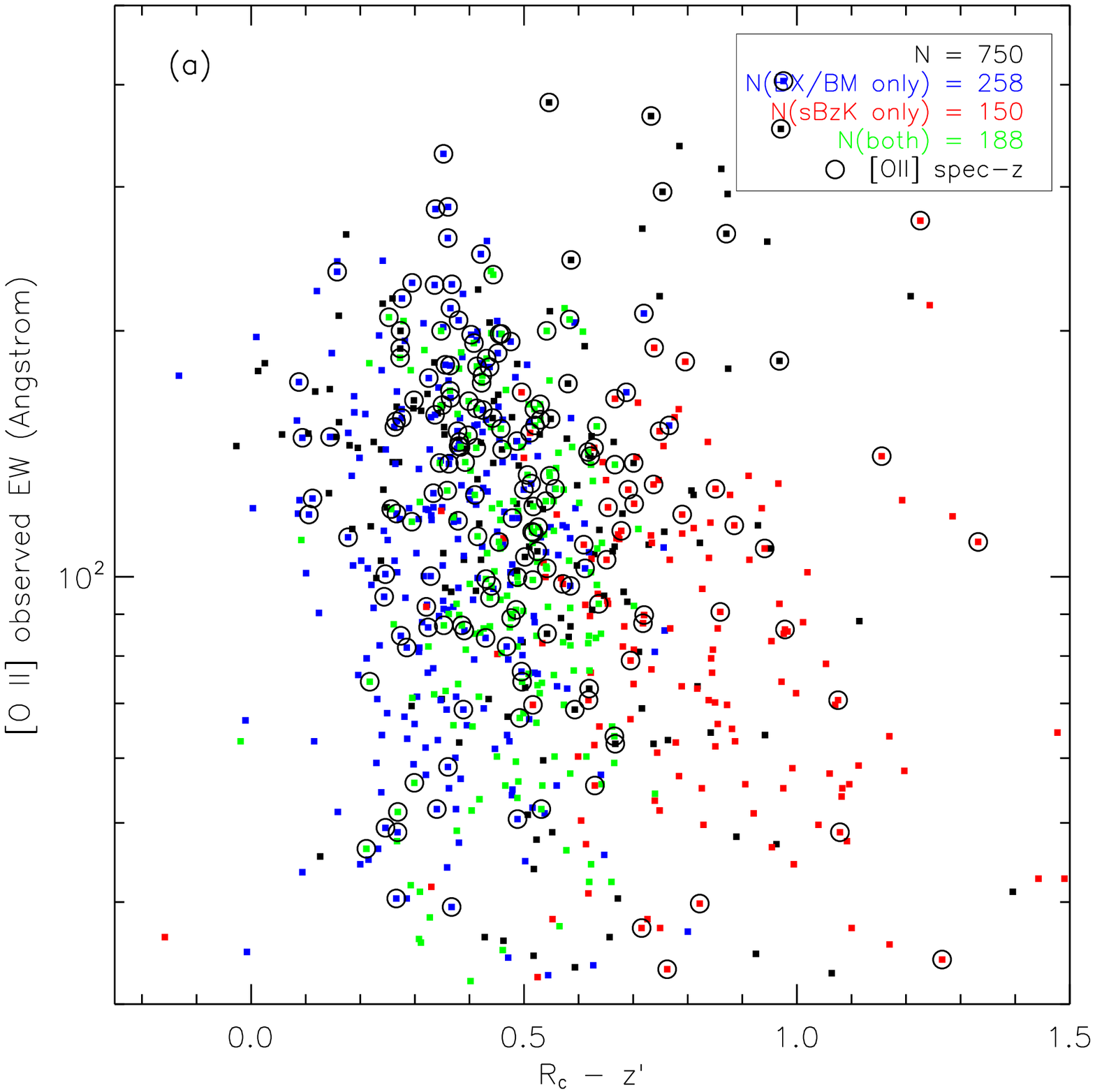}{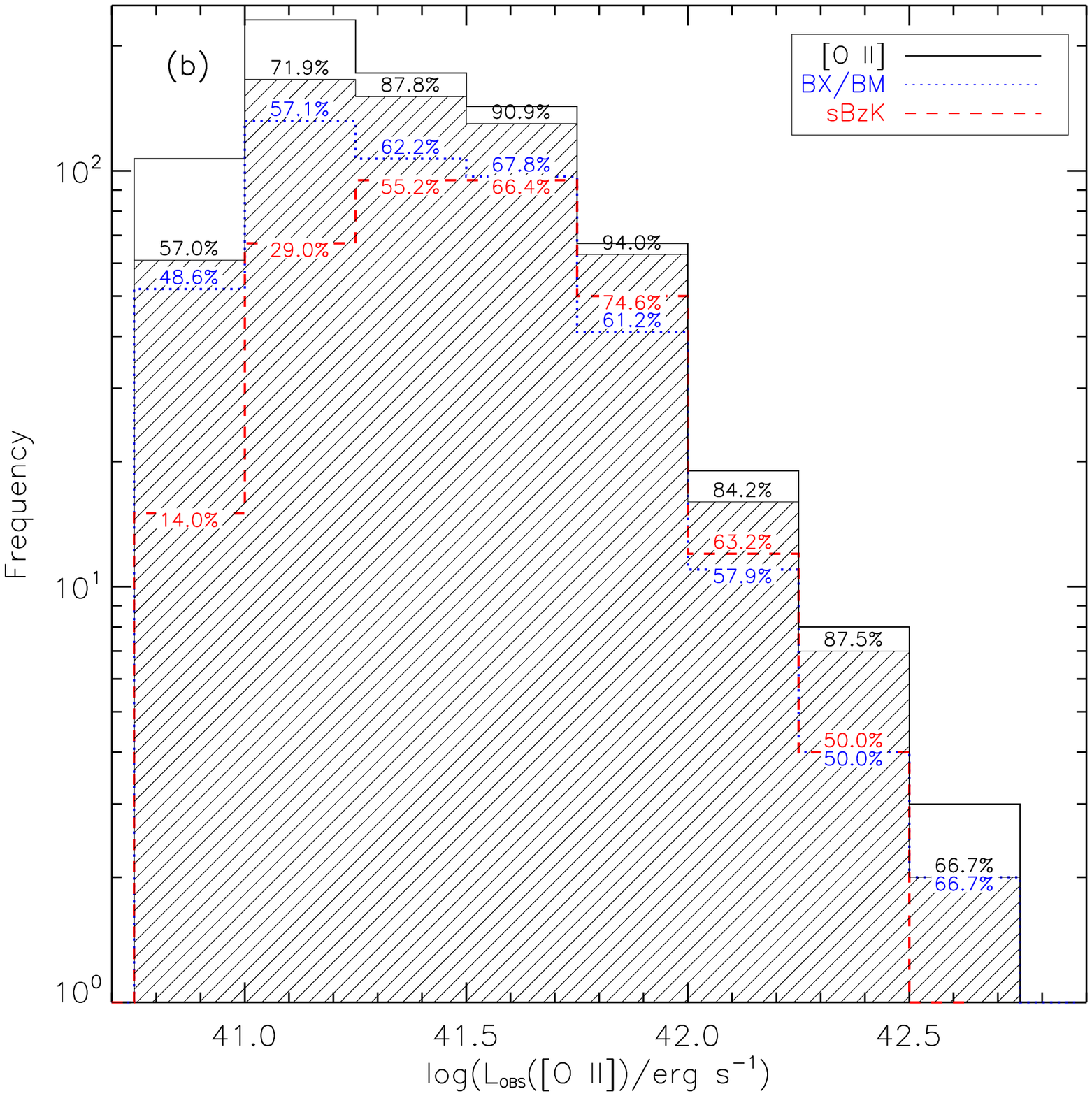}
  \caption{({\it a}) \OII\ observed EW as a function of \Rc--$z$\arcmin\ color
    for NB921 emitters. \OII\ emitters that are selected purely as
    ``BX/BM'' ($\Rcf<26.0$ mag) and sBzK are shown as blue and red squares,
    respectively. Those identified using both color selection techniques are
    shown as green squares, and those that do not meet both selections are
    shown as black squares. Spectroscopically confirmed \OII\ emitters are
    shown as black circles. ({\it b}) The probability distribution of
    \OII\ observed luminosity, as shown by the solid black line. The red
    dashed line illustrates the distribution for \OII\ sBzK galaxies.
    Likewise, the blue dotted line shows the \OII\ BX/BM galaxies.
    The combination of both method is shown by the shaded
    histogram. Percentages are reported relative to the full
    \OII\ sample.}
  \label{OII_color}
\end{figure*}

We illustrate in Figure~\ref{OII_color}(a) the observed
\OII\ emission-line EW as a function of \Rc--$z$\arcmin.
It is clear that the BX/BM and sBzK techniques are complementary
by identifying the bluest and reddest galaxies, respectively.
Roughly 25\% of the \OII\ emitters are identified using both color
selections, while 34\% and 20\% are identified by only the BX/BM
and sBzK selections, respectively.
Figure~\ref{OII_color}(a) also shows that $\sim$20\% (154/750)
of the \OII\ sample is not identified by either method.
These galaxies are generally of lower luminosity where 43\% of
them are fainter than $\Rcf=26.0$ mag and 84\% are undetected
(at 3$\sigma$) in $K$. These galaxies have typical
limits on their stellar mass of $\sim3\times10^8$ \Msun.
We note that the \OII\ emitters galaxies with $\Rcf<26.0$ but
which still miss the BX/BM selection have typical stellar masses
of $2\times10^9$ \Msun, which is 0.25 dex more (0.86 less) massive
than the BX/BM (sBzK). This suggests that a subset of
intermediate mass star-forming galaxies are missed by color selections.

Finally the distribution of \OII\ luminosity for the sBzK
and BX/BM samples are shown in Figure~\ref{OII_color}(b) where it
illustrates that either of these color selections will miss
40\%--50\% of the \OII\ sample. Even combining both methods,
roughly 20\% of the faintest \OII\ emitters are missed.

%
%
\section{Conclusions}\label{sec:5}
Using deep NB imaging with Suprime-Cam, we identified \nOII\ \OII\ emitting
galaxies at $z$=1.5--1.6. Follow-up optical and near-infrared spectroscopy
confirmed that the two-color selection of \OII\ emitters is highly reliable:
Only 1\% of NB921 \OII\ emitters appeared to be missed while 3\% of $z=0.84$
\OIII\ emitters contaminate the \OII\ NB921 color selection.
Our survey has investigated the properties of \OII-emitting galaxies. We find:
\begin{enumerate}
  \setlength{\itemsep}{1pt}
  \setlength{\parskip}{0pt}
  \setlength{\parsep}{0pt}
  \item The average rest-frame \OII\ EW of our $z=\za$ sample is 45\AA,
    consistent with another study at $z>0.7$. The EW distribution
    declines gradually from the peak towards lower EWs, suggesting
    that very few $z\sim1.5$ galaxies lack \OII\ emission.
  \item We compared the observed \OII\ and UV luminosities and find
    a strong correlation with $\log{L(\textsc{O ii})} \propto (\SFRslope) \log{L(1500)}$.
    The observed scatter (0.23 dex) of individual \OII\ measurements
    against the UV luminosity suggests that the \OII\ luminosity can be
    considered as a coarse SFR indicator.
  \item We constructed multi-band SEDs (rest-frame 1500\AA--1.8\mm),
    and find that higher EW galaxies have slightly bluer colors
    compared to lower EW galaxies.
  \item We fitted the SEDs with exponentially declining ``$\tau$''
    models, and find a median stellar ages of 100 Myr with 13\%
    of the sample older than 500 Myr. The stellar masses are
    typically $10^{9.5\pm0.6}$ \Msun.
  \item We find that \OII\ emission is present in the full census of
    galaxies, from massive red systems (found by sBzK) to smaller,
    star-bursting systems (found by BX/BM). With the sensitivity
    that we achieved in the NB921 filter, we find that 154 ($\approx$20\%)
    $z=\za$ \OII\ emitters are not identified with BzK or BX/BM
    because they are too faint (i.e., stellar mass limits of
    $\sim3\times10^8$ \Msun).
    A subset of these un-identified galaxies that are bright enough
    to satisfy the BX/BM magnitude limit ($\Rcf=26.0$) have
    intermediate stellar masses ($2\times10^9$ \Msun) compared to
    the BX/BM and sBzK samples, and suggest that some intermediate
    mass galaxies are currently missed by commonly used high-$z$
    color selections.
\end{enumerate}

\acknowledgements
We thank Nelson Caldwell with help on designing Hectospec fiber
configurations, Richard Cool for additional help using the
MMT/Hectospec \textsc{hsred} reduction pipeline, and Ryosuke Goto
for assistance in the FMOS observations.
The authors wish to recognize and acknowledge the very significant
cultural role that the summit of Mauna Kea has always
had within the indigenous Hawaiian community.
We are most fortunate to have the opportunity to conduct
observations from this mountain.

{\it Facilities:} \facility{Subaru (Suprime-Cam, FOCAS, FMOS)},
\facility{MMT (Hectospec)}, \facility{Keck (DEIMOS)},
\facility{{\it GALEX}}, \facility{Mayall (MOSAIC, NEWFIRM)},
\facility{UKIRT (WFCAM)}, \facility{Spitzer (IRAC)}

\end{document}